\documentclass[twocolumn,english,aps,prb,floatfix,amssymb,superscriptaddress,longbibliography]{revtex4}

\usepackage{amsmath} 
\usepackage{amssymb} 
\usepackage{amsthm} 
\usepackage{amsfonts} 
\usepackage{listings} 
\usepackage{longtable} 
\usepackage{enumerate} 
\usepackage{latexsym} 
\usepackage{color} 
\usepackage{setspace} 
\usepackage{blindtext} 
\usepackage{dsfont} 
\usepackage{chemformula} 
\usepackage{mathrsfs} 
\usepackage{xcolor} \definecolor{ggr}{gray}{0.95}

\usepackage{array,etoolbox} \preto\tabular{ 
\setcounter{magicrownumbers}{0}} 
\newcounter{magicrownumbers}

\usepackage{bm} 
\usepackage{hyperref}

\usepackage{psfrag}

\usepackage{bm} 
\usepackage{graphicx} 
\usepackage{subfigure}

\RequirePackage[normalem]{ulem} 

\RequirePackage{color}\definecolor{RED}{rgb}{1,0,0}\definecolor{BLUE}{rgb}{0,0,1} 

\usepackage{siunitx} \DeclareSIUnit\angstrom{\text {Å}}

\newcommand{\beq}{ 
\begin{equation}
} 
\newcommand{\eneq}{ 
\end{equation}
} 
 
\renewcommand{\vec}[1]{\ensuremath{\boldsymbol{#1}}}

\newcommand{\iu}{\mathrm{i}}

\newcommand{\pdag}{{\phantom{\dagger}}} 
\newcommand{\qsl}{\mathrm{QSL}} 
\renewcommand{\top}{\mathrm{t}} 
\renewcommand{\bot}{\mathrm{b}} 
\newcommand{\tun}{\mathrm{tb}}

\newcommand{\bra}[1]{\left\langle#1\right|} 
\newcommand{\ket}[1]{\left|#1\right\rangle}

\input{epsf}

\begin{document} 

\title{Probing quantum spin liquids with a quantum twisting microscope}

\author{Valerio Peri} 
\email{vperi@caltech.edu} \affiliation{ Department of Physics, IQIM, California Institute of Technology, Pasadena, CA 91125, USA. }

\author{Shahal Ilani} \affiliation{ Department of Condensed Matter Physics, Weizmann Institute of Science, Rehovot 76100, Israel. }

\author{Patrick A. Lee} \affiliation{ Department of Physics, Massachusetts Institute of Technology, Cambridge, MA 02139, USA. }

\author{Gil Refael} \affiliation{ Department of Physics, IQIM, California Institute of Technology, Pasadena, CA 91125, USA. } 
\begin{abstract}
The experimental characterization of quantum spin liquids poses significant challenges due to the absence of long-range magnetic order, even at absolute zero temperature. The identification of these states of matter often relies on the analysis of their excitations. In this paper, we propose a method for detecting the signatures of the fractionalized excitations in quantum spin liquids using a tunneling spectroscopy setup. Inspired by the recent development of the quantum twisting microscope, we consider a planar tunneling junction, in which a candidate quantum spin liquid material is placed between two graphene layers. By tuning the relative twist angle and voltage bias between the leads, we can extract the dynamical spin structure factor of the tunneling barrier with momentum and energy resolution. Our proposal presents a promising tool for experimentally characterizing quantum spin liquids in two-dimensional materials. 
\end{abstract}

\date{\today}

\maketitle

\section{Introduction} Quantum spin liquids (QSLs) are states of matter that defy magnetic ordering even at temperatures far below their exchange energy. While theoretical understanding of QSLs has made significant progress \cite{Wen:2002,Zhou:2017,Savary:2017}, experimental verification of these states of matter remains a formidable challenge \cite{Knolle:2019,Wen:2019,Broholm:2020}. The absence of order serves as an imperfect definition, as it is impossible to rule out all possible ordered states. A more experimentally accessible characteristic of QSLs is the presence of emergent gauge field and low-energy excitations with fractionalized quantum numbers, such as spinons, which carry fractional spin and zero charge.

Several experimental techniques have been employed to investigate the signatures of fractionalized excitations in potential QSL materials. Thermal transport measurements, for instance, have been used to explore the low-energy physics of $\alpha$-\ch{RuCl3} \cite{Leahy:2017}, volborthites \cite{Watanabe:2016}, $\kappa$-\ch{(ET) 2Cu2 (CN)3} \cite{Yamashita:2009}, and organic dmits \cite{Yamashita:2010}. Optical absorption and Raman spectroscopy have provided access to the zero-momentum excitation spectrum of pyrochlores \cite{Maczka:2008}, herbertsmithite \cite{Wulferding:2010}, and Kitaev materials \cite{Little:2017,Wang:2017,Wellm:2018,Sandilands:2015}. Nuclear magnetic resonance and muon spin relaxation have enabled to locally probe these candidate materials \cite{Yaouanc:2010,Carretta:2011}. Additionally, spin transport \cite{Chatterjee:2015,Chen:2013}, NV centers \cite{Chatterjee:2019,Khoo:2022,Lee:2023}, and numerous other techniques \cite{Mross:2011,Morampudi:2017,Chen:2020,Norman:2009,Aftergood:2020,Mazzilli:2023} have been proposed for investigating QSLs. Inelastic neutron scattering \cite{Han:2012,Lake:2013,Mourigal:2013,Banerjee:2017,Paddison:2017,Banerjee:2016}, however, stands out due to its unique advantage of probing the excitation spectrum with both momentum and energy resolution. Unfortunately, a major limitation of inelastic neutron scattering is its reliance on large bulk three-dimensional crystals. This drawback renders it unsuitable for studying QSLs in the mono- and few-layer limit, which attracted considerable interest with the emergence of high-quality two-dimensional materials \cite{Novoselov:2005}.

\begin{figure}
\includegraphics{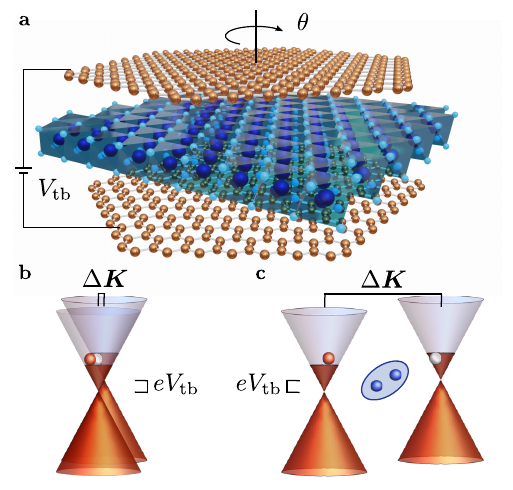} 
\caption{{\bf a} Schematic representation of the tunneling heterostructure. The top and bottom layers are graphene, while the middle barrier is a monolayer of a candidate QSL material, e.g., 1T-\ch{TaS2}. $V_\tun$ is the voltage bias between the graphene layers, while $\theta$ is the relative twist angle between them. Two distinct tunneling process can take place. {\bf b} Direct tunneling can occur only if $\hbar v_\mathrm{F}\lvert\Delta\vec{K}\rvert < eV_\tun+2\mu$ such that the momentum and energy of the graphene's electron are conserved. The tunneling process corresponds to a particle-hole excitation across the layers. {\bf c} The low energy excitation of the barrier can mediate additional inelastic tunneling processes through which the graphene's electron changes its momentum and energy. The electron can then tunnel even when the Dirac cones of the two layers are further apart compared to {\bf b}, i.e., $\hbar v_\mathrm{F}\lvert\Delta\vec{K}\rvert>eV_\tun+2\mu$. An example of a QSL excitations, i.e., a spinon particle-hole excitation, is here schematically represented in blue. \label{fig1}} 
\end{figure}

Inspired by the recent development of the quantum twisting microscope \cite{Inbar:2023}, we consider a planar junction formed by two graphene layers separated by a QSL material, depicted in Fig.~\ref{fig1} {\bf a}, as a technique for investigating the fractionalized excitations in two-dimensional materials. Traditional tunneling probes offer limited momentum resolution, thereby restricting their capability to fully characterize these excitations \cite{Fernandez-Rossier:2009,Fransson:2010,Klein:2018,Ghazaryan:2018,Carrega:2020, Konig:2020, Feldmeier:2020,Chen:2022,Ruan:2021,He:2022,He:2023a,He:2023,Jia:2022,Kao:2023,Mitra:2023,Eickhoff:2020,Bauer:2023}. In our proposal, the momentum resolution is achieved by controlling the relative twist angle $\theta$ between the graphene layers. A finite twist angle introduces a momentum mismatch $\Delta\vec{K}$ between the Dirac cones of the two graphene layers. In the absence of the QSL barrier, as depicted in Fig.\ref{fig1} {\bf b}, the direct tunneling of electrons can occur if $\hbar v_\mathrm{F}\lvert\Delta\vec{K}\rvert < eV_\tun+2\mu$, where $eV_\tun$ is the voltage bias between the layers and $\mu$ the chemical potential \cite{Bistritzer:2010,Mishchenko:2014,Guerrero-Becerra:2016,Inbar:2023}. If we continue increasing the twist angle while keeping the voltage bias fixed, the direct tunneling gets suppressed. When the tunneling barrier hosts low-energy excitations, however, additional processes come into play. The graphene's electrons can change their energy and momentum by scattering from the excitations in the QSL. Consequently, these additional inelastic processes contribute to the total current even when $\hbar v_\mathrm{F}\lvert\Delta\vec{K}\rvert>eV_\tun+2\mu$ \cite{Fernandez-Rossier:2009,Fransson:2010,Klein:2018} (see Fig.~\ref{fig1} {\bf c}). This inelastic contribution to the tunneling current offers valuable insights into the excitation spectrum and the underlying dynamics of the QSL. 

The central result of our study is captured by Eq.~\eqref{eq:golden}. By considering the second derivative of the tunneling current with respect to the voltage bias, i.e, the inelastic electron tunneling spectroscopy (IETS) signal, we access the dynamical spin structure factor of the tunneling barrier at momentum $\Delta \vec{K}$ and energy $eV_\tun$. The twist angle determines the separation between the Dirac cones of the two graphene layers and controlling it provides the sought-after momentum resolution. This additional tunable knob enables qualitative and quantitative distinction among various QSLs, facilitating the inference of the microscopic Hamiltonian governing a candidate QSL material. 

The development of probes capable of providing momentum and energy resolution similar to inelastic neutron scattering but tailored for two-dimensional materials is highly desirable. This is particularly important considering the potential existence of materials that exhibit QSL behavior exclusively in the monolayer limit. One intriguing example is 1T-\ch{TaS2}, a layered transition metal dichalcogenide (TMD). Below a critical temperature of approximately \SI{200}{\kelvin}, it undergoes a commensurate $\sqrt{13}\times\sqrt{13}$ charge-density wave (CDW) transition. The CDW arrangement forms a triangular lattice of Stars of David, each containing an odd number of electrons per unit cell. The residual Coulomb interaction in this system induces a Mott-insulating gap of \SI{400}{\milli\electronvolt}\cite{Wilson:1974,Fazekas:1979}. Interestingly, no magnetic ordering has been observed down to very low temperatures, leading to suggestions that it may host a gapless QSL \cite{Law:2017,He:2018,Ribak:2017,Manas-Valero:2021,Yu:2017}. However, in the bulk material, the formation of interlayer dimers can compete with the Mott physics picture, potentially opening a trivial band-insulator gap \cite{Ritschel:2018,Wang:2020,Martino:2020,Butler:2020,Lee:2021}. Nevertheless, the possibility of observing a QSL in a monolayer of 1T-\ch{TaS2} remains promising. 

In this work, our primary focus is on QSLs, which are intriguing and elusive states of matter that demand new experimental probes. Nonetheless, our proposal equally applies to the study of arbitrary magnetic barriers. For instance, it can be used to investigate magnon excitations in magnetically ordered materials \cite{Klein:2018,Ghazaryan:2018,Mitra:2023}. Therefore, we anticipate that the quantum twisting microscope \cite{Inbar:2023} will expand our understanding of magnetic interactions in two-dimensional materials.

The remainder of the manuscript is organized as follows. In Sec.~\ref{sec:setup}, we introduce the setup, describing the tunneling through a QSL in a quantum twisting microscope. In Sec.~\ref{sec:analytical}, we demonstrate how, under a set of simplifying assumptions, the inelastic electron tunneling spectroscopy signal contains a contribution directly proportional to the dynamical spin structure factor of the QSL. Sec.~\ref{sec:quantum spin liquid} focuses on studying the spin structure factor of different types of QSLs via a mean-field approximation and showcases how our proposal can distinguish among them. Finally, we provide concluding remarks in Section~\ref{sec:conclude}.

\section{Tunneling through a quantum spin liquid in a quantum twisting microscope \label{sec:setup} 
} We investigate the vertical tunneling of electrons between two twisted graphene layers, separated by a QSL material that acts as a tunneling barrier. The twist angle between the graphene layers, denoted by $\theta$, can be controlled in situ, and a voltage bias, denoted by $V_\tun$, is applied across them. While a similar setup has been explored in Ref.~\onlinecite{Carrega:2020}, the tunability of the twist angle was not considered. The recent development of the quantum twisting microscope \cite{Inbar:2023} motivates us to investigate the potential of this additional degree of freedom. 

The Hamiltonian describing the junction is given by: 
\begin{equation}
H=\underbrace{H_\top+H_\bot+H_\qsl}_{H_0}+H_\tun, 
\end{equation}
where $H_\top$ and $H_\bot$ are the Hamiltonians of the top and bottom graphene layers, respectively, $H_\qsl$ characterizes the QSL serving as a tunneling barrier, and $H_\tun$ is the tunneling Hamiltonian. 

The specific form of $H_\qsl$, which determines the low-energy excitations responsible for the inelastic electron scattering during tunneling, is not crucial for our derivations and will be discussed in Sec.~\ref{sec:quantum spin liquid}. This flexibility makes our proposal suitable for exploring a wide range of magnetic materials beyond QSLs.

We model each graphene layer as a gas of massless Dirac particles around the valley $\vec{K}_\ell$, where $\ell=\{\top,\bot\}$ labels the layer. This approximation holds for energies up to a few hundred \si{\milli\electronvolt} and for disorder and interaction strengths that do not induce significant inter-valley scattering. We will discuss at the end of this section how to correctly account for electrons in the vicinity of the $\vec{K}'_\ell$ valley. The Hamiltonian of the $\ell$ layer is: 
\begin{equation}\label{eq:graphene1layer} 
H_\ell = \hbar v_\mathrm{F}\sum_{\vec{k},\alpha,\beta,s}\left[\vec{\tau}_{\alpha\beta} \cdot\left( \vec{k}+\vec{K}_\ell\right)-\mu_\ell\right]\,c^\dagger_{\vec{k},s,\alpha,\ell}c^{\phantom{\dagger}}_{\vec{k},s,\beta,\ell} \,, 
\end{equation}
where $c^\dagger_{\vec{k},s,\alpha,\ell}$ creates an electron in sublattice $\alpha=\{\mathrm{A},\mathrm{B}\}$ with momentum $\vec{k}$ and spin $s=\{\uparrow,\downarrow\}$. The Pauli matrices $\vec{\tau}$ act on the sublattice degree of freedom, $v_\mathrm{F}\approx \SI{e6}{\meter\per\second}$ is the Fermi velocity of graphene, and $\mu_\ell$ is the chemical potential of layer $\ell$ measured from the charge neutrality point of the layer. In Eq.~\eqref{eq:graphene1layer}, the momentum $\vec{k}$ in each layer is measured with respect to the Dirac point $\vec{K}_\ell$. Due to the twist angle $\theta$, $\vec{K}_\bot$ and $\vec{K}_\top$ do not coincide, i.e., $\Delta\vec{K}=\vec{K}_\bot-\vec{K}_\top=\vec{K}_\bot-R(\theta)\vec{K}_\bot$, where $R(\theta)$ is the matrix implementing a rotation by an angle $\theta$. 

The real-space tunneling Hamiltonian is given by \cite{Carrega:2020}: 
\begin{equation}\label{eq:tunnellRS} 
\begin{split}
	H_\tun&=\frac{1}{\sqrt{N}}\sum_{\vec{R}_\top,\vec{R}_\bot}\sum_{\alpha,\beta,s,s'}\Gamma_{ss'}(\Delta \vec{r})\\&\quad\times c^\dagger_{\vec{R}_\top,\alpha,s,\top}c^{\phantom{\dagger}}_{\vec{R}_\bot,\beta,s',\bot}e^{\mathrm{i}eV_\tun t/\hbar}+\mathrm{H.c.}, 
\end{split}
\end{equation}
where $\vec{R}_\top$ and $\vec{R}_\bot$ are the unit cell position in the top and bottom layer, respectively, and the voltage bias across the junction is incorporated as a time-dependent hopping process. The tunneling matrix $\Gamma_{ss'}(\Delta \vec{r})$ is a function of the in-plane distance between the initial and final positions of the electron, denoted as $\Delta\vec{r} = (\vec{R}_\top+\vec{r}_{\top,\alpha})-(\vec{R}_\bot+\vec{r}_{\bot,\beta})$. Here, $\vec{r}_{\alpha\ell}$ represents the location of the sublattice $\alpha$ within the unit cell of layer $\ell$. $\Gamma_{ss'}(\Delta \vec{r})$ consists of a bare tunneling term, where the electron does not interact with the QSL, and a term describing tunneling via an exchange-mediated excitation of the QSL: 
\begin{equation}\label{eq:tunnelMatrix} 
\Gamma_{ss'}(\Delta \vec{r})=\Gamma_0(\Delta \vec{r})\,\delta_{s,s'}+\Gamma_1(\Delta \vec{r})\,\vec{\sigma}_{ss'}\cdot \vec{s}(\vec{r}_m). 
\end{equation}
The Pauli matrices $\vec{\sigma}$ act on the spin degree of freedom of the electrons in the leads, and $\vec{s}(\vec{r}_m)$ represents the local magnetic moment in the insulating barrier at $\vec{r}_m=(\vec{R}_\top+\vec{r}_{\top,\alpha}+\vec{R}_\bot+\vec{r}_{\bot,\beta})/2$. This tunneling matrix provide an accurate description when the exchange energy $\approx J_\mathrm{g} \lvert\vec{s}\rvert$ is smaller than the spin-independent barrier height $\Phi$. In this limit, we have $\Gamma_1/\Gamma_0\approx J_\mathrm{g}/\Phi$, with $J_\mathrm{g}$ the exchange coupling between the graphene's electrons and the magnetic moments of the QSL \cite{Fransson:2010,Fernandez-Rossier:2009}. 

In Eq.~\eqref{eq:tunnelMatrix}, we assumed that the scattering from the localized spins occurs at the midpoint $\vec{r}_m$, neglecting the microscopic details of the barrier. This assumption simultaneously maximizes the transition amplitudes from the top and bottom graphene layers to the magnetic barrier \cite{Carrega:2020}. Additionally, we neglected Kondo interactions between the localized moments and the conducting electrons of each layer. This assumption is justified in the limit of low density in the leads and strong magnetic correlations in the QSL. Lastly, we did not include the effect of RKKY interactions, which is a good approximation when the QSL's lattice constant $a_\mathrm{QSL}$ is larger than the graphene's one $a_\mathrm{g}=\SI{2.46}{\angstrom}$. In this case, these interactions are random in sign and weak compared to the magnetic correlations. 

We can Fourier transform Eq.~\eqref{eq:tunnellRS} and obtain: 
\begin{equation}\label{eq:FTAll} 
\begin{split}
	H_\tun&=\frac{1}{\sqrt{N}}\sum_{\vec{k},\vec{k}'}\sum_{\vec{g}_\top,\vec{g}_\bot}\sum_{\alpha,\beta, s,s'}e^{\mathrm{i}\left(\vec{g}_\top\cdot \vec{r}_{\alpha\top}-\vec{g}_\bot\cdot \vec{r}_{\beta\bot}\right)} \\
	&\quad\times[\Gamma_0(\vec{q}')\delta_{s,s'}\delta_{\vec{q}}+\Gamma_1(\vec{q}')\vec{\sigma}_{ss'}\cdot \vec{s}_{\vec{q}}]\\&\quad\times c^\dagger_{\vec{k},\alpha, s,\top}c^{\phantom{\dagger}}_{\vec{k}',\beta, s',\bot}e^{\mathrm{i}eV_\tun t/\hbar}+\mathrm{H.c.}, 
\end{split}
\end{equation}
with $\vec{q}'=\left(\vec{k}+\vec{k}'+\vec{K}_\top+\vec{K}_\bot+\vec{g}_\top+\vec{g}_\bot\right)/2$, and $\vec{q}=\vec{k}-\vec{k}'+\vec{K}_\top-\vec{K}_\bot+\vec{g}_\top-\vec{g}_\bot$. Here, $\vec{g}_\top$ and $\vec{g}_\bot$ are the reciprocal lattice vectors of top and bottom graphene layer, respectively. Note that the momenta $\vec{q}$ and $\vec{q}'$ are measured from the center of the Brillouin zone $\vec{\Gamma}$. Instead, $\vec{k}$ and $\vec{k}'$ are measured from $\vec{K}_\top$ and $\vec{K}_\bot$, respectively. As in Eq.~\eqref{eq:graphene1layer}, we consider the tunneling of electrons in the vicinity of the $\vec{K}$ valleys and hence assume small $\vec{k}$ and $\vec{k}'$. At the end of this section, we will comment on how to correctly account for the tunneling of electrons around the $\vec{K}'$ valley. 

As in the Bistritzer-MacDonald model for twisted bilayer graphene \cite{Bistritzer:2011}, the tunneling functions $\Gamma_0(\vec{q}')$ and $\Gamma_1(\vec{q}')$ rapidly decay as a function of $\lvert \vec{q}' \rvert$, since the vertical separation of the leads exceeds the in-plane lattice constant \cite{Bistritzer:2011, Carrega:2020}. We thus consider only the components of the tunneling matrix near the Dirac points of the unrotated graphene layers, i.e., $\lvert\vec{q}'\rvert\approx\left\lvert\vec{K}_\top+\vec{K}_\bot\right\rvert/2$, and denote the momentum-independent tunneling amplitudes as $\bar{\Gamma}_0$ and $\bar{\Gamma}_1$. This simplification constraints the sum over reciprocal lattice vectors in Eq.~\eqref{eq:FTAll} and we obtain: 
\begin{equation}\label{eq:tunnelingFT} 
\begin{split}
	H_\tun&=\frac{1}{\sqrt{N}}\sum_{\vec{k},\vec{q}}\sum_{\alpha,\beta,s,s'}\sum_{n=0}^2 \left[\bar{\Gamma}_0\delta_{s,s'}\delta_{\vec{q},\Delta\vec{K}_n}+\bar{\Gamma}_1\vec{\sigma}_{ss'}\cdot \vec{s}_{\vec{q}_n}\right] \\
	&\quad\times T^{(n)}_{\alpha\beta} \Big(c^\dagger_{\vec{k},\alpha,s,\top}c^{\phantom{\dagger}}_{\vec{k}-\vec{q},\beta,s',\bot}e^{\mathrm{i}eV_\tun t/\hbar} \\&\quad\quad\quad\quad\quad+ c^\dagger_{\vec{k}-\vec{q},\alpha,s,\bot}c^{\phantom{\dagger}}_{\vec{k},\beta,s',\top}e^{-\mathrm{i}eV_\tun t/\hbar}\Big), 
\end{split}
\end{equation}
with $\vec{q}_n=\vec{q}+\Delta\vec{K}_n$, and $\Delta\vec{K}_n$ denotes the separation of the Dirac cones of the top and bottom layers at the three $\vec{K}$ valleys in the Brillouin zone. These three vectors are related by a \SI{120}{\degree} rotation, i.e., $\Delta\vec{K}_n=R(2\pi n/3)\Delta\vec{K}$. The tunneling matrices are: 
\begin{equation}\label{eq:grapheneTN} 
T^{(n)}=\tau^0+\cos\left(\frac{2\pi n}{3}\right)\tau^x-\sin\left(\frac{2\pi n}{3}\right)\tau^y. 
\end{equation}

For a sufficiently large barrier height, the tunneling Hamiltonian $H_\tun$ acts as a perturbation and the current flowing through the heterostructure can be computed using linear response theory: 
\begin{equation}\label{eq:currentLinear} 
I(eV_\tun) = \frac{\mathrm{i}}{\hbar}\int_{-\infty}^{+\infty}\mathrm{d}t'\theta(t-t')\left\langle\left[H_\tun(t'),I_\tun(t)\right]\right\rangle. 
\end{equation}
The current operator in Eq.~\eqref{eq:currentLinear} is given by: 
\begin{equation}\label{eq:currentComm} 
\begin{split}
	I_\tun &= \frac{\mathrm{i}e}{\hbar}\left[H_\tun,N_\top\right]\\&= -\frac{\mathrm{i}e}{\hbar\sqrt{N}} \sum_{\vec{k},\vec{q}}\sum_{\alpha,\beta,s,s'}\sum_{n=0}^2 \left[\bar{\Gamma}_0\delta_{s,s'}\delta_{\vec{q},\Delta\vec{K}_n}+\bar{\Gamma}_1\vec{\sigma}_{ss'}\cdot \vec{s}_{\vec{q}_n}\right]\\&\quad\times T_{\alpha\beta}^{(n)}\Big(c^\dagger_{\vec{k},\alpha,s,\top}c^{\phantom{\dagger}}_{\vec{k}-\vec{q},\beta,s',\bot}e^{\mathrm{i}eV_\tun t/\hbar}\\& \quad\quad\quad\quad\quad-c^\dagger_{\vec{k}-\vec{q},\alpha,s,\bot}c^{\phantom{\dagger}}_{\vec{k},\beta,s',\top}e^{-\mathrm{i}eV_\tun t/\hbar}\Big). 
\end{split}
\end{equation}

The current consists of two distinct contributions: direct tunneling, which leaves the magnetic moments of the barrier unperturbed, and a spin-flip process proportional to the spin fluctuations of the QSL. The direct tunneling contribution is given by (see App.~\ref{app:one}): 
\begin{equation}\label{eq:directTun} 
\begin{split}
	I^{(0)}(eV_\tun)&=\frac{4 e \bar{\Gamma}_0^2}{\hbar}\sum_{n=0}^2 \mathcal{A}^\tun(\Delta\vec{K}_n,eV_\tun,n). 
\end{split}
\end{equation}
The spin-dependent part, instead, is (see App.~\ref{app:one}): 
\begin{equation}\label{eq:final} 
\begin{split}
	I^{(2)}(eV_\tun) &= -\frac{2 e \bar{\Gamma}_1^2}{\hbar}\int_{-\infty}^{+\infty} \mathrm{d}\omega\left[n_{\mathrm{B}}(\omega)- n_{\mathrm{B}}(\omega+eV_\tun)\right] \\
	&\quad\times\sum_n\sum_{\vec{q}}\,\mathcal{S}(\vec{q}_n,eV_\tun+\omega)\,\mathcal{A}^\tun(\vec{q},-\omega,n), 
\end{split}
\end{equation}
where $n_\mathrm{B}$ is the Bose-Einstein distribution. The excitations of the QSL are described by the dynamical spin structure factor $\mathcal{S}(\vec{q},\omega)=\sum_\gamma\mathcal{S}^\gamma(\vec{q},\omega)$ defined as $\mathcal{S}^\gamma(\vec{q},\omega) \stackrel{\iu\omega\to\omega+\mathrm{i}\epsilon^+}{=}-\frac{1 }{\pi}\mathrm{Im} \mathcal{G}_\mathrm{s}^\gamma(\vec{q},\iu \omega_n)$, with 
\begin{equation}\label{eq:spinonSpinStructure} 
\begin{split}
	\mathcal{G}_\mathrm{s}^\gamma(\vec{q},\iu \omega_n)&=-\int_0^\beta\mathrm{d}\tau e^{\iu\tau\omega_n}\sum_{ij}e^{\iu\vec{q}\cdot(\vec{r}_i-\vec{r}_j)}\langle \mathcal{T}s^\gamma_{i}(\tau)s^\gamma_{j}\rangle. 
\end{split}
\end{equation}
Here, $\omega_n$ are bosonic ($=2n\pi/\beta$) Matsubara frequencies. We will further characterize the spin structure factor in Sec.~\ref{sec:quantum spin liquid}.

In Eqs.~\eqref{eq:directTun} and \eqref{eq:final}, we introduced the spectral function $\mathcal{A}^\tun$ which describes the particle-hole excitations in the graphene layers and is defined as: 
\begin{equation}\label{eq:chiTB} 
\begin{split}
	\mathcal{A}^\tun(\vec{q},\omega,n)&=\frac{\pi}{N}\sum_{\vec{k},\lambda,\lambda'}\int_{-\infty}^{+\infty}\mathrm{d}\epsilon\left[n_{\mathrm{F}}(\epsilon)-n_{\mathrm{F}}(\epsilon+\omega)\right]\\
	&\quad\times\left\lvert T^{(n)}_{\vec{k},\vec{k}-\vec{q};\lambda,\lambda'} \right\rvert^2\mathcal{A}^\top_{\lambda}(\vec{k},\epsilon)\mathcal{A}^\bot_{\lambda'}(\vec{k}-\vec{q},\epsilon+\omega). 
\end{split}
\end{equation}
Here, $\mathcal{A}_\lambda^\ell(\vec{k},\omega)$ represents the spectral function of band $\lambda$ in the graphene layer $\ell$, and $n_\mathrm{F}$ is the Fermi-Dirac distribution. The tunneling matrix $ T ^{(n)}_{\vec{k},\vec{k}-\vec{q};\lambda,\lambda'}$ is obtained by projecting Equation \eqref{eq:grapheneTN} onto the eigenbasis of the graphene layers: 
\begin{equation}\label{eq:projectedT} 
\begin{split}
	T ^{(n)}_{\vec{k},\vec{k}-\vec{q};\lambda,\lambda'}&= \langle\psi_{\lambda}(\vec{k})\vert T^{(n)}\vert\psi_{\lambda'}(\vec{k}-\vec{q})\rangle\\
	&=\frac{1}{2}\left[1+\lambda e^{\iu\left(\frac{2\pi n}{3}+\phi_{\vec{k}}\right)}\right]\left[1+\lambda' e^{-\iu\left(\frac{2\pi n}{3}+\phi_{\vec{k}-\vec{q}}\right)}\right], 
\end{split}
\end{equation}
where $\lvert\psi_\lambda (\vec{k})\rangle=(\lambda e^{-\mathrm{i}\phi_{\vec{k}}},1)^\mathrm{T}/\sqrt{2}$ represents the eigenvector of Equation \eqref{eq:graphene1layer} in band $\lambda$ at momentum $\vec{k}=\lvert\vec{k}\rvert e^{\mathrm{i}\phi_{\vec{k}}}$.

While the graphene layers are $C_{6z}$-symmetric, the candidate QSL material may lack this symmetry. This consideration requires to be careful when computing the additional contribution to the inelastic tunneling stemming from the electrons in the vicinity of the $\vec{K}'$ valleys. The shift of the Dirac cones at the $\vec{K}'$ valley is opposite to that at the $\vec{K}$ one, i.e., $\Delta\vec{K}'_n=-\Delta\vec{K}_n$. Therefore, to account for the contribution from the opposite valley, it is not sufficient to multiply by a valley degeneracy factor $g_v=2$. Instead, we have to substitute $\mathcal{S}(\vec{q}+\Delta\vec{K}_n)$ in Eq.~\eqref{eq:final} with $\mathcal{S}(\vec{q}-\Delta\vec{K}_n)$.

In general, we expect an additional contribution to the total current that is linear in the magnetic moment of the barrier $\langle \vec{s}_{\vec{q}_n} \rangle$. However, this term vanishes when the leads are not spin-polarized (see App.~\ref{app:one}) \cite{Fransson:2010}. Our primary objective is to investigate the properties of QSLs, with a particular emphasis on measuring the spin structure factor $\mathcal{S}(\vec{q},\omega)$ rather than $\langle \vec{s}_{\vec{q}_n} \rangle$. The absence of an additional current term linear in $\bar{\Gamma}_1$ is advantageous for our purposes. Nevertheless, the use of spin-polarized leads would offer the opportunity to probe individual components of $\mathcal{S}(\vec{q},\omega)$ \cite{Franke:2022}.

\section{Inelastic electron tunneling spectroscopy and the spin structure factor \label{sec:analytical} 
} Eqs.~\eqref{eq:directTun} and \eqref{eq:final} provide a general description of the current flowing through the heterostructures, accounting for finite temperature, disorder, and interactions in the leads. Our objective is to establish a simple analytical relation that directly connects the dynamical spin structure factor $\mathcal{S}(\vec{q},\omega)$ with a measurable quantity in the tunneling heterostructure. To achieve this, we make a series of simplifying approximations.

To distinguish the features of fractionalized excitations from the thermal broadening of magnons in magnetically ordered materials \cite{Franke:2022}, one has to consider a regime where the thermal energy $k_\mathrm{B}T$ is much smaller than the characteristic magnetic correlations $J$ within the tunneling barrier, i.e., $k_\mathrm{B}T\ll J$. Therefore, we consider the limit of zero temperature in the subsequent analysis. The expression for the spin-dependent current simplifies as follows: 
\begin{equation}
\begin{split}\label{eq:partialI2} 
	I^{(2)}_{T=0}(eV_\tun) &= \frac{2 e \bar{\Gamma}_1^2}{\hbar}\int_{0}^{eV_\tun} \mathrm{d}\omega\sum_n\sum_{\vec{q}}\,\mathcal{S}(\vec{q}_n,\omega) \\
	&\quad\times\mathcal{A}^\tun(\vec{q},eV_\tun-\omega,n), 
\end{split}
\end{equation}
with 
\begin{equation}
\begin{split}
	\mathcal{A}^\tun_{T=0}(\vec{q},\omega,n)&=\frac{\pi}{N}\sum_{\vec{k},\lambda,\lambda'}\int_{0}^{\omega}\mathrm{d}\epsilon\left\lvert T^{(n)}_{\vec{k},\vec{k}-\vec{q};\lambda,\lambda'} \right\rvert^2\\
	&\quad\times\mathcal{A}^\top_{\lambda}(\vec{k},\epsilon-\omega)\mathcal{A}^\bot_{\lambda'}(\vec{k}-\vec{q},\epsilon). 
\end{split}
\end{equation}
The current is a convolution of the spin structure factor $\mathcal{S}(\vec{q}_n,\omega)$ and the particle-hole spectral function $\mathcal{A}^\tun_{T=0}(\vec{q},\omega,n)$ which is a property of the graphene layers. This convolution acts as a frequency-dependent smearing of the momentum at which we probe $\mathcal{S}$.

Let us consider electron-doped graphene layers and further assume $\mu_\top=\mu_\bot\gg eV_\tun$. This situation can be achieved by independently controlling the bottom and top gates of the tunneling junction. The largest momentum transfer $\lvert\vec{q} \rvert$ is then $\lvert \vec{q}_\mathrm{max}\rvert\approx 2\mu/\hbar v_\mathrm{F}$. If we focus on sufficiently large twist angles, i.e., $2\mu/\hbar v_\mathrm{F}\lvert\Delta\vec{K}\rvert\ll 1$, we can disregard the smearing in Eq.~\eqref{eq:partialI2}. Namely, we ignore the momentum transfer $\vec{q}$ between the graphene layers and the QSL, and carry the sum over $\vec{q}$ past $\mathcal{S}(\vec{q}_n,\omega)$. 

The scattering matrix $T^{(n)}_{\vec{k},\vec{k}-\vec{q}}$ depends exclusively on the angles $\phi_{\vec{k}}$ and $\phi_{\vec{k}-\vec{q}}$, cf. Eq.~\eqref{eq:projectedT}, and is unity once averaged over the angular variables. We can then use $1/N\sum_{\vec{k}} \mathcal{A}(\vec{k},\epsilon)=\mathcal{N}(\epsilon)$, where $\mathcal{N}(\epsilon)=\sqrt{3}a_\mathrm{g}^2\lvert \mu+\epsilon\rvert/(4\pi\hbar^2 v_\mathrm{F}^2)$ is the particle density of the graphene layer. In the limit $\mu \gg eV_\tun$, we neglect the energy dependence of the graphene's density of state and reach a simple expression for the spin-dependent tunneling current: 
\begin{equation}\label{eq:ietsSpin} 
\begin{split}
	I^{(2)}_{T=0}(eV_\tun)&= -\frac{3 e a^2_\mathrm{g}L^2\bar{\Gamma}^2_1\mu^2}{ 8\pi\hbar^5v_\mathrm{F}^4}\sum_n\\
	&\quad\times\int_0^{eV_\tun}\mathrm{d}\omega\mathcal{S}(\Delta\vec{ K}_n,\omega)(eV_\tun-\omega), 
\end{split}
\end{equation}
where $L$ is the in-plane linear size of the graphene layers.

The scattering from the excitations of the QSL opens new inelastic channels for the electrons tunneling between the graphene layers. This should result in a clear signature in the inelastic electron tunneling spectroscopy. Taking the second derivative of Eq.~\eqref{eq:ietsSpin}, we obtain the spin-dependent contribution to the IETS signal: 
\begin{equation}\label{eq:golden} 
\frac{\mathrm{d}^2I^{(2)}_{T=0}(eV_\tun)}{\mathrm{d}V_\tun^2}=-\frac{3 e^3 a^2_\mathrm{g}L^2\bar{\Gamma}^2_1\mu^2}{ 8\pi\hbar^5v_\mathrm{F}^4}\sum_n\mathcal{S}(\Delta \vec{K}_n,eV_\tun). 
\end{equation}
By measuring the IETS signal in a quantum twisting microscope with a magnetic insulator acting as the tunneling barrier, it becomes possible to directly probe the spectrum of magnetic excitations. The control of the voltage bias between the graphene layers and the relative twist angle between them gives access to the energy and momentum dependence of the spin structure factor, respectively.

While Eq.~\eqref{eq:final} is a generic result that relies on few physically-motivated assumptions and constitutes a rigorous starting point for a comparison with future experimental data, Eq.~\eqref{eq:golden} is a simple analytical relation resting on a series of additional assumptions. Specifically, it holds at zero temperature, for large twist angles, and for voltage biases much smaller than the graphene's chemical potential. These limits capture the regime where the IETS signal can be directly linked to the spin structure factor of the barrier, as described by Eq.~\eqref{eq:golden}.

In App~.\ref{grapheneeh}, we compute the current stemming from direct tunneling and show that direct tunneling also contributes to the IETS signal \cite{Inbar:2023}. The spin-independent contribution, however, is only present for voltages simultaneously satisfying $eV_\tun <\hbar v_\mathrm{F}\lvert\Delta\vec{K}\rvert< eV_\tun+2\mu $. Considering that the relevant energy scale of the QSL is expected to be on the order of tens of \si{\milli\electronvolt}, the large Fermi velocity of graphene ensures that a significant region of the $eV_\tun-\theta$ diagram exclusively contains the IETS signal originating from the QSL contribution. This is clearly depicted in Fig.~\ref{fig2}, where the gray region shades the area where the direct tunneling contributes. Note that it is also the region where the simplifying assumptions made to reach Eq.~\eqref{eq:golden} break down.

The energy resolution of the quantum twisting microscope is set by the measurement temperature. Preliminary results have been obtained with a quantum twisting microscope operating at \SI{4}{\kelvin} \cite{Inbar:2023a,Birkbeck:2023}. While this temperature is already significantly lower than the magnetic interactions in many candidate QSL materials, reaching even lower temperatures will be crucial to conclusively probe the excitations of these materials. The momentum resolution on $\Delta \vec{K}$ depends on the ability to control the twist angle of the leads. Previous studies \cite{Inbar:2023} achieved a resolution of $\delta\theta\approx\SI{0.001}{\degree}$, which corresponds to $\delta\lvert\Delta\vec{K}\rvert \approx \SI{3e-5}{\per\angstrom}$. The main uncertainty in probing the spin-structure factor at a definite momentum, however, stems from precisely accounting for the convolution with the graphene's particle-hole spectral function in Eq.~\eqref{eq:partialI2}. There are two main factors that limit the momentum resolution. First, the in-plane linear size $L$ of the graphene layers sets a limit of $\delta\lvert\vec{q}\rvert \approx 1/L$. Second, via the convolution in Eq.~\eqref{eq:partialI2}, the signal is smeared over the Fermi surface of the graphene layers with a resolution $\delta \lvert\vec{q}\rvert \approx 2\mu/\hbar v_\mathrm{F}$. It would appear that the best resolution is achieved when both graphene layers are maintained at charge neutrality. Nonetheless, in this configuration, one cannot neglect the linear dependence in energy of the graphene's density of state. The energy convolution in Eq.~\eqref{eq:ietsSpin} would then be with the cubic rather than linear power of $(eV_\tun-\omega)$, and only the fourth derivative of the current with respect to the voltage bias would be directly proportional to the spin structure factor. Instead, one could use a two-dimensional metal with a small Fermi pocket at valley $\vec{K}$ and achieve similar results to those discussed here for doped graphene. Lastly, we neglected any spatial inhomogeneity in the graphene layers, that can in principle modulate the tunneling amplitude on a length scale $\ell$. If present, they would further reduce the momentum resolution by roughly $1/\ell$. 

The maximum momentum that can be probed is approximately $\lvert \Delta\vec{K} \rvert_\mathrm{max} = 2 \lvert\vec{K}\rvert\sin\theta_\mathrm{max}/2$, with previous studies reaching up to $\theta_\mathrm{max}=\SI{30}{\degree}$ \cite{Birkbeck:2023,Inbar:2023a}. Given that the unit cell of graphene is often smaller than that of candidate QSL materials, even small twist angles should be able to explore a significant portion of the QSL Brillouin zone. For example, for 1T-\ch{TaS2} we have $a_\mathrm{QSL}\approx\SI{12.11}{\angstrom}$ and an angle $\theta\approx\SI{11.7}{\degree}$ suffices to probe the corners of the BZ. By tuning the twist angle between the graphene layers, it is possible to probe line-cuts in the Brillouin zone of the candidate QSL material. To fully map the angular dependence of the spin structure factor, however, it would be desirable to also control the angle between graphene and the QSL barrier.

\section{Distinguishing various quantum spin liquids \label{sec:quantum spin liquid} 
} To analyze different types of QSLs, we adopt a phenomenological approach by using a slave-particle mean-field Hamiltonian \cite{Baskaran:1988,Affleck:1988,Dagotto:1988,Wen:1996,Senthil:2000,Wen:2002a}. While this approach may not capture the precise details of specific materials, its main purpose is to demonstrate the efficacy of a quantum twisting microscope in distinguishing between various QSLs. Importantly, the key result represented by Eq.~\eqref{eq:golden} remains valid regardless of the specific details of the QSL Hamiltonian or the methods employed to compute the spin structure factor $\mathcal{S}(\vec{q},\omega)$.

We consider a tunneling barrier described by a Hamiltonian quadratic in the spin degree of freedom, such as a Heisenberg Hamiltonian. To describe the physical spins, we utilize a fermionic representation in terms of Abrikosov fermions and decouple the quartic term in the spinon operators via a mean-field approximation.
%
The dynamical spin structure factor can be directly computed using the definition of Eq.~\eqref{eq:spinonSpinStructure} (see App.~\ref{app:three}). For instance, in the case of a $U(1)$ QSL, it is given by: 
\begin{equation}\label{eq:finalquantum spin liquids} 
\begin{split}
	\mathcal{S}(\vec{q},\omega) &= \frac{1}{4N} \sum_{ss'}\sum_{\vec{k}}\sum_{nl} g_{ss'}(\vec{k},\vec{q},n,l) (2-\delta_{ss'})\\&\quad\times\left[n_\mathrm{F}(\xi^s_{n\vec{k}})-n_\mathrm{F}(\xi^{s'}_{l\vec{k}+\vec{q}})\right]\delta\left(\omega-\xi^s_{n\vec{k}}+\xi^{s'}_{l\vec{k}+\vec{q}}\right), 
\end{split}
\end{equation}
where $\xi^s_{n\vec{k}}$ is the energy at momentum $\vec{k}$ of the band $n$ with spin $s$ of the spinon mean-field Hamiltonian. The associated eigenvector at sublattice $\alpha$ is $U_{\alpha ns}(\vec{k})$. The form factor $g_{ss'}(\vec{k},\vec{q},n,l)$ accounts for the overlap of the eigenstates' wavefunctions: 
\begin{equation}
g_{ss'}(\vec{k},\vec{q},n,l)=\left\vert\sum_{\alpha} e^{\mathrm{i}\vec{q}\cdot\vec{\delta}_\alpha}U^*_{\alpha ns}(\vec{k})U_{\alpha ls'}(\vec{k}+\vec{q}) \right\vert^2, 
\end{equation}
where $\vec{\delta}_\alpha$ is the position of sublattice $\alpha$ within the unit cell. The form factor plays an important role in multi-band spinon models, whereas we will neglect it in the low-energy and single-band models analyzed below.

The dynamical spin structure factor describes spinon particle-hole excitations rather that a single quasiparticle with a well-defined dispersion, as it would be for magnons in magnetically ordered materials. In fact, spin-flip excitations must have integer spin while spinons have fractional spin. Consequently, the energy and momentum of the excitation are shared between two quasiparticles, leading to a broad continuum of excitations rather than a sharp dispersion. This broad continuum in the excitation spectrum is characteristic of QSL behavior.

Now, let us consider three qualitatively distinct quantum spin liquids: a Dirac spin liquid, a gapped chiral QSL, and a gapless QSL with a spinon Fermi surface. For each case, we derive the dynamical spin structure factor $\mathcal{S}(\vec{q},\omega)$ in the long-wavelength and low-energy limit at zero temperature \cite{Chatterjee:2015}. 

\begin{figure}
\includegraphics{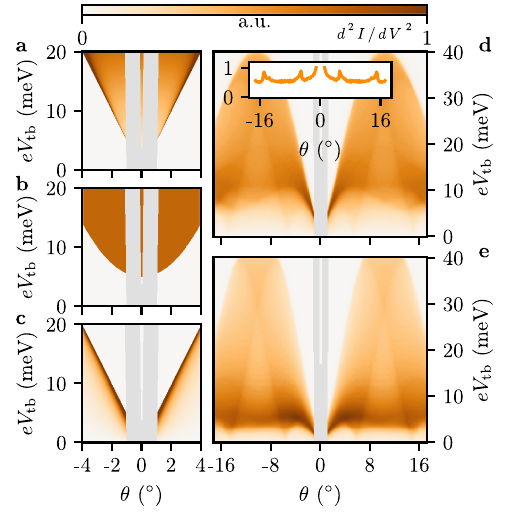} 
\caption{Inelastic electron tunneling spectroscopy signal for various QSLs in a quantum twisting microscope: {\bf a} a Dirac QSL, {\bf b} a gapped QSL, and {\bf c} a QSL with a spinon Fermi surface with a parabolic spinon band. For the evaluation of the IETS signal, we assumed $v_\mathrm{s}= \SI{2.5e4}{\meter\per\second}$, $\Delta_\mathrm{s}=\SI{30}{\milli\electronvolt}$, $m_\mathrm{s}\approx 3 m_\mathrm{e}$, and $\mu=\SI{100}{\milli\electronvolt}$. {\bf d} and {\bf e} present the IETS signal of a QSL with a spinon Fermi surface obtained by a spinon tight-binding model on a triangular lattice with and without next-to-nearest neighbor hoppings, respectively. The inset in {\bf d} shows the IETS signal at zero bias voltage. Here, we assumed $t_\mathrm{s}=\SI{5}{\milli\electronvolt}$ and $\chi_\mathrm{s}=\SI{1}{\milli\electronvolt}$ for the spinon hopping. The gray region shades the area where elastic tunneling contributes to the IETS signal.\label{fig2}} 
\end{figure}

Let us start with the Dirac spin liquid. Its low-energy dispersion is given by $\xi_{n\vec{k}}=n\hbar v_\mathrm{s}\lvert\vec{k}\rvert$, with $n=\pm$ and $v_\mathrm{s}$ the spinon Fermi velocity. The dynamical spin structure factor for this case is obtained as: 
\begin{equation}\label{eq:kinematicquantum spin liquid2} 
\begin{split}
	\mathcal{S}(\vec{q},\omega)& = \frac{3\Omega}{2}\int \frac{\mathrm{d}^2\vec{k}}{(2\pi)^2}\delta\left(\omega-\hbar v_\mathrm{s}\lvert\vec{k}\rvert-\hbar v_\mathrm{s}\lvert\vec{k}+\vec{q}\rvert\right)\\&= \frac{3\Omega}{16\pi\hbar^2 v_\mathrm{s}^2}\frac{\omega^2-\hbar^2 v_\mathrm{s}^2 q^2/2}{\sqrt{\omega^2-\hbar^2 v_\mathrm{s}^2 q^2}}\Theta\left(\omega-\hbar v_\mathrm{s}q\right), 
\end{split}
\end{equation}
where $\Omega$ is the unit-cell area of the QSL. Note that this expression equally applies to the case of $\mathbb{Z}_2$ Dirac spin liquids. The spin structure factor given by Eq.~\eqref{eq:kinematicquantum spin liquid2} exhibits a square-root singularity as $\omega$ approaches $\hbar v_\mathrm{s}q^+$ and a threshold at $\omega> \hbar v_\mathrm{s}q$, which is determined by the spinon Fermi velocity, cf. Fig~\ref{fig2} {\bf a}. 

In the case of gapped chiral QSLs, we assume that the spinon dispersion has a minimum at $\vec{k}=0$ for the conduction band and a maximum for the valence band at the same location. The low-energy dispersion is given by $\xi_{n\vec{k}}=n\left(\hbar^2\vec{k}^2/2m_\mathrm{s}+\Delta_\mathrm{s}/2\right)$, where $n=\pm$. $\Delta_\mathrm{s}$ is the gap in the spinon spectrum and $m_\mathrm{s}$ is the spinon mass. The dynamical spin structure factor is obtained as: 
\begin{equation}\label{eq:kinematicquantum spin liquid3} 
\begin{split}
	\mathcal{S}(\vec{q},\omega) &= \frac{3\Omega}{2}\int \frac{\mathrm{d}^2\vec{k}}{(2\pi)^2}\delta\left(\omega-\Delta_\mathrm{s}-\frac{\hbar^2 k^2}{2m_\mathrm{s}}-\frac{\hbar^2 (\vec{k}+\vec{q})^2}{2m_\mathrm{s}}\right) \\
	& = \frac{3m_\mathrm{s}\Omega}{8\pi\hbar^2}\Theta\left(\omega-\Delta_\mathrm{s}-\frac{\hbar^2 q^2}{4m_\mathrm{s}}\right). 
\end{split}
\end{equation}
The result holds for gapped $\mathbb{Z}_2$ QSLs as well. The spin structure factor described by Eq.~\eqref{eq:kinematicquantum spin liquid3} exhibits a step-like behavior with a threshold at $\omega > \Delta_\mathrm{s}+\hbar^2q^2/4m_\mathrm{s}$, as depicted in Fig.~\ref{fig2} {\bf b}.

Finally, let us consider a gapless QSL with a spinon Fermi surface. We approximate the spinon dispersion with a parabola and consider the limit $\lvert\vec{q} \rvert \ll k_\mathrm{s,F}$, where $k_\mathrm{s,F}$ is the spinon Fermi momentum. The spin structure factor is given by \cite{Nozieres:1999}: 
\begin{equation}\label{eq:kinematicquantum spin liquid} 
\begin{split}
	\mathcal{S}(\vec{q},\omega) &= \frac{3}{2}\frac{1}{2 k_\mathrm{s,F}q_n}\int_{\substack{k<k_\mathrm{s,F}\\
	\lvert \vec{k}+\vec{q}\rvert>k_\mathrm{s,F}}} \frac{\mathrm{d}^2\vec{k}}{(2\pi)^2}\\&\quad\times\delta\left(\omega+\frac{\hbar^2k^2}{2m_\mathrm{s}}-\frac{\hbar^2(\vec{k}+\vec{q})^2}{2m_\mathrm{s}}\right) \\
	& =\frac{3}{8\pi^2 \hbar qv_\mathrm{s}}\frac{\omega}{\sqrt{\hbar^2 v_\mathrm{s}^2q^2-\omega^2}}\Theta\left(\hbar v_\mathrm{s}q-\omega\right), 
\end{split}
\end{equation}
where $v_\mathrm{s}=\hbar k_\mathrm{s,F}/m_\mathrm{s}$ and $m_\mathrm{s}$ is the effective mass of the spinons. The spin structure factor, shown in Fig.~\ref{fig2} {\bf c}, decreases linearly at small frequencies and has a finite signal only for $\omega<\hbar v_\mathrm{s}q$.

By analyzing the IETS signal in the quantum twisting microscope, we can qualitatively distinguish between various QSLs. Since the direct tunneling dominates at small twist angle and large voltage bias, it is beneficial to control the twist angle to access the large-angle regime where the inelastic contribution prevails at low voltage biases, as shown in Fig.~\ref{fig2}. Examining how the IETS signal evolves at low applied bias voltage allows to differentiate among different quantum spin liquids, as discussed in the previous paragraphs \cite{Chatterjee:2015, Carrega:2020}. While Ref.~\onlinecite{Konig:2020} first showed the possibility to single out the IETS signal of a gapped QSL in a planar tunneling junction lacking the twist angle tuning knob, at odds with previous proposals \cite{Konig:2020,Carrega:2020, Konig:2020}, the additional momentum resolution provided by the twist angle allows to also extract low-energy parameters of the spinon Hamiltonian, such as the effective mass and the Fermi velocity of the spinons.
 
To further prove how the momentum resolution offers insights into the underlying Hamiltonian of the QSL, let us consider a QSL with a spinon Fermi surface on a triangular lattice. This state has been suggested to be a good description for the putative QSL in 1T-\ch{TaS2} \cite{He:2018}. At the mean-field level, this state can be described by a nearest-neighbor tight-binding model for spinons on the triangular lattice \cite{Li:2017a}. The spinon energy is spin-independent and given by $\xi_{\vec{k}}=-2t_\mathrm{s}\Big[\cos{(k_x)}+2\cos{(\frac{k_x}{2})}\cos{(\frac{\sqrt{3} k_y}{2})}\Big]-\mu_\mathrm{s}$, where $t_\mathrm{s}$ is the spinon hopping parameter, and the chemical potential $\mu_\mathrm{s}=0.8346\,t_\mathrm{s}$ ensures half-filling of the spinon bands. We consider a spinon hopping $t_\mathrm{s}=\SI{5}{\milli\electronvolt}$. Note that the actual value for 1T-\ch{TaS2} is currently unknown. Attempts to fit the residual temperature-independent static spin susceptibility and specific heat measured in 1T-\ch{TaS2} \cite{Klanjsek:2017,Ribak:2017} to a simple mean-field spinon model \cite{He:2018} result in an order of magnitude larger spinon hopping parameter. These measurements, however, were carried out in bulk samples and it is unclear whether they should be linked to a QSL model \cite{Ritschel:2018,Wang:2020,Martino:2020,Butler:2020}. STM studies reporting evidence of QSL behavior in the closely related 1T-\ch{TaSe2} monolayer, instead, put an upper bound of \SI{5}{\milli\electronvolt} on the in-plane exchange coupling \cite{Ruan:2021,Chen:2022}. While at the present time we cannot reliably estimate the physical value of $t_s$, we stress that our proposal works best for QSLs with a small exchange coupling. Namely, a larger separation in the QSL's and graphene's energy scales improves the validity of the approximations that led to Eq.~\eqref{eq:golden} and increases the area in the $eV_\tun-\theta$ diagram where the IETS signal is dominated by the QSL contribution. Note also that the spinon spectral signatures takes place at voltage biases smaller than the charge gap, e.g., \SI{400}{\milli\electronvolt} for 1T-\ch{TaS2}. Therefore, spinon excitations can be easily separated from charge excitations. The IETS signal arising from this state, as obtained from Eq.\eqref{eq:finalquantum spin liquids} and Eq.\eqref{eq:golden}, is shown in Fig.~\ref{fig2} {\bf d}. The inset shows the IETS signal at zero voltage bias. It proves how the IETS signal as a function of the twist angle can determine the spinon Fermi momentum. In fact, the peaks of the IETS signal at zero voltage bias and non-zero twist angle correspond to $2\lvert\vec{K}\rvert\sin(\theta/2)=\lvert 2\vec{k}_{\mathrm{F},s}+\vec{b}_i\rvert$, where $\vec{b}_i$ is a QSL reciprocal lattice vector.  

An additional next-nearest-neighbor hopping term for the spinons represents a minimal modification to the previous model \cite{Li:2017a}. The new energy dispersion is given by $\xi'_{\vec{k}}=\xi_{\vec{k}}-2\chi_\mathrm{s}\Big[\cos{(\sqrt{3}k_y)}+2\cos{(\frac{3k_x}{2})}\cos{(\frac{\sqrt{3} k_y}{2})}\Big]$, and the chemical potential is adjusted to maintain half-filling. The resulting IETS signal is shown in Fig.~\ref{fig2} {\bf e}. We observe a significant shift of weight towards lower voltage biases. While this example is particularly simple, it illustrates how the experimentally measurable IETS signal, when compared to different microscopic theories, contains sufficient information to better constrain the Hamiltonian of the underlying QSL material. Ultimately, the finite resolution may limit the ability to precisely determine the microscopic parameters, but the possibility to map the spin structure factor in both momentum and energy space presents a substantial advantage over existing probes for QSL materials in mono- and few-layer materials.

In Fig.~\ref{fig2}, we presented the IETS signal expected in a quantum twisting microscope according to Eq.\eqref{eq:golden}. These results have been obtained fixing the chemical potential of the graphene layers to $\mu=\SI{100}{\milli\electronvolt}$, which meets the condition $\mu\gg eV_\tun$ for the voltage biases considered. Nonetheless, such a large value of the chemical potential carries unwanted consequences. To reach Eq.~\eqref{eq:golden} from Eq.~\eqref{eq:partialI2}, we neglected the momentum transfer to the QSL $\vec{q}$, whose maximum value is $\delta\vec{q}=2\mu/\hbar v_\mathrm{F}\approx \SI{3e-2}{\per\angstrom}$, or alternatively $\delta\theta\approx\SI{0.7}{\degree}$.
Here, we show how relaxing the condition $\mu\gg eV_\tun$ to $\mu\geq eV_\tun$ allows to reduce the uncertainty on momentum and still distinguish among various QSLs.

\begin{figure}
\includegraphics{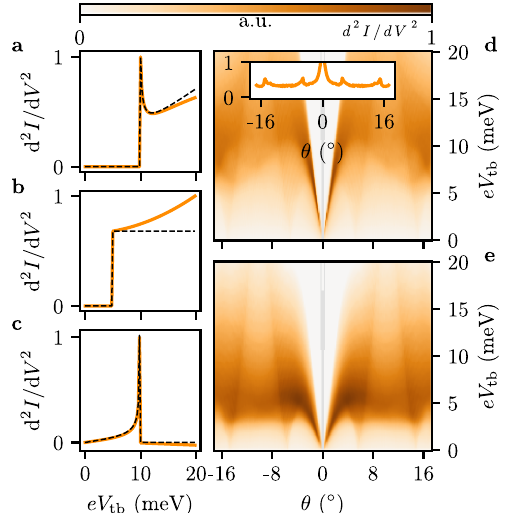} 
\caption{Inelastic electron tunneling spectroscopy signal with smaller chemical potential: {\bf a} a Dirac QSL, {\bf b} a gapped QSL, and {\bf c} a QSL with a spinon Fermi surface with a parabolic spinon band. For the evaluation of the IETS signal, we assumed the same parameters as in Fig.~\ref{fig2} but $\mu=\SI{20}{\milli\electronvolt}$ and fixed the angle at $\theta=\SI{3}{degrees}$. The black dashed line indicates the IETS signal obtained via Eq.~\eqref{eq:golden}, while the orange line the one obtained by Eq.~\eqref{eq:smallermu} {\bf d} and {\bf e} are analogous to the same panels in Fig.~\ref{fig2} but with IETS signal computed according to Eq.~\eqref{eq:smallermu} and $\mu=\SI{20}{\milli\electronvolt}$.\label{fig3}} 
\end{figure}

By requiring $\mu\geq eV_\tun$, we can still restrict our attention to scattering involving only electrons in the upper band, but we cannot ignore the linear-in-energy density of state of graphene. In this case, we obtain:
\begin{equation}
	\label{eq:smallermu}
	\begin{split}
	\frac{\mathrm{d}^2I^{(2)}_{T=0}(eV_\tun)}{\mathrm{d}V_\tun^2}=&-\frac{3 e^3 a^2_\mathrm{g}L^2\bar{\Gamma}^2_1}{ 8\pi\hbar^5v_\mathrm{F}^4}\sum_n\Bigg[\mu^2\mathcal{S}(\Delta \vec{K}_n,eV_\tun)\\&-\int_0^{eV_\tun}\mathrm{d}\,\omega\mathcal{S}(\Delta \vec{K}_n,\omega)(eV_\tun-\omega)\Bigg]\,.
	\end{split}
	\end{equation}
In Fig.~\ref{fig3}, we compare the IETS signal at a fixed angle $\theta=\SI{2}{\degree}$ obtained by Eq.~\eqref{eq:golden} and by Eq.~\eqref{eq:smallermu} with $\mu=\SI{20}{\electronvolt}$. 
The chemical potential $\mu=\SI{20}{\milli\electronvolt}$ corresponds to an error $\delta\vec{q}\approx\SI{6e-3}{\per\angstrom}$. 
In Fig.~\ref{fig3} {\bf a}, we present the result for Dirac QSL, where we can see that while at large voltages the signal differs, we still detect an onset at $eV_\tun=\hbar v_\mathrm{s}\lvert\Delta \vec{K}\rvert$ with a square-root singularity above it, cf. Eq.~\eqref{eq:kinematicquantum spin liquid2}. 
A similar analysis is displayed in Fig.~\ref{fig3} {\bf b} for a gapped chiral QSL. We again observe an onset at $eV_\tun=\Delta_\mathrm{s}+\hbar^2\lvert\Delta \vec{K}\rvert^2/2m_\mathrm{s}$, as expected from Eq.~\eqref{eq:kinematicquantum spin liquid3}, but the signal is no longer constant above it. 
The absence of a singularity at the onset allows to distinguish this case from that of a Dirac QSL. Lastly, in Fig.~\ref{fig3} {\bf c}, we present the IETS signal of a QSL with a spinon Fermi surface. 
Here, the signal ceases to be zero above the threshold $eV_\tun=\hbar v_\mathrm{s}\lvert\Delta \vec{K}\rvert$. The presence of a finite signal below the threshold with a linear dependence on $eV_\tun$ at small voltage biases and a square-root singularity when the threshold is approached from below clearly distinguishes this case from the previous ones. In Fig.~\ref{fig3} {\bf d} and {\bf e}, we show that even with Eq.~\eqref{eq:smallermu} and a smaller chemical potential, it is possible to identify the spinon Fermi wavevector and the shift of the IETS signal at lower voltage biases upon the introduction of next-to-nearest neighbor hopping in toy model of spinon hopping on a triangular lattice.
	
Lastly, we note that for QSL with large energy scales, even the condition $\mu\geq V_\tun$ might result in too large $\delta\vec{q}$. In this case, we foresee two possible strategies. On one hand, we could change the chemical potential as a function of the voltage bias such that at low energies, the IETS signal is still well described by Eq.~\eqref{eq:golden} with a small momentum uncertainty. On the other hand, one could numerically solve Eq.~\eqref{eq:final} in its full generality to compare various theoretical models to the experimental results. 

\section{Discussions and Conclusions \label{sec:conclude} 
} In this manuscript, we introduced the quantum twisting microscope as a promising tool for imaging the excitations of QSLs in two-dimensional materials. One significant advantage of this technique, distinguishing it from previous proposals, is its ability to provide resolution in both momentum and energy. By focusing on simplified limits, we demonstrated that the IETS signal of the tunneling junction has a contribution directly proportional to the dynamical spin structure factor of the tunneling barrier, as shown in Eq.~\eqref{eq:golden}. 

Refs.~\onlinecite{Konig:2020,Carrega:2020} first proposed to use a planar tunneling junctions to probe quantum spin liquids. Ref.~\onlinecite{Konig:2020} showed how to effectively separate between the elastic and inelastic contribution to the tunneling current in this configuration. In their proposal, however, only the zero-momentum spin structure factor of quantum spin liquids with short-range spinon correlations could be probed. Using graphene as metallic leads and introducing a fixed finite twist angle, Ref.~\onlinecite{Carrega:2020} showed how to probe the spin structure factor of a Kitaev QSL at finite momentum. Nonetheless, it did not investigate the possibility to continuously vary the twist angle or the signal stemming from different QSLs. Here, motivated by these previous works and the quantum twisting microscope of Ref.~\onlinecite{Inbar:2023}, we showed how the in-situ control over the twist angle is an important tuning knob that allows to probe the spin structure factor with energy and momentum resolution. Conceptually, our proposal provides a tool to study candidate QSL materials in the mono- and few-layer limit in ways that were previously achievable only for bulk materials via inelastic neutron scattering. The additional momentum resolution will allow to extract precious information on the effective parameters of the microscopic Hamiltonian describing the QSL, e.g., the spinon velocity, the Fermi wavevectors, and the spinon mass. 

In the following, we briefly discuss the implications of departing from the limits considered in our study and the experimental feasibility of our proposal. Eq.~\eqref{eq:final} is valid at finite temperatures, and in the presence of disorder and interaction in the leads. Moreover, it holds regardless of the approximation used to compute the spin structure factor. While not as simple as Eq.~\eqref{eq:golden}, it establishes a starting point for comparing experimental results to theoretical predictions. Additionally, a faithful representation of the experimental conditions requires a careful description of the electrostatics of the junction. In this study, we assumed that the voltage bias through the junction results in a simple electrostatic potential shift. However, in reality, a voltage bias will induce both a shift of the chemical potential and an electrostatic potential between the layers. This more realistic situation can be easily accounted for \cite{Inbar:2023,Mishchenko:2014,Guerrero-Becerra:2016}. A dual-gated device, with independent control of the gates of the bottom and top graphene layers \cite{Inbar:2023}, offers a more direct realization of the present proposal as it allows to maintain fixed chemical potentials and change the relative electrostatic potential shift.

From a fabrication standpoint, substantial progress has been made in the mechanical exfoliation of mono- and few-layer samples of candidate quantum spin liquids such as $\alpha$-\ch{RuCl3} and 1T-\ch{TaS2} \cite{Du:2018,Zhou:2019,Mashhadi:2018,Yu:2015,Tsen:2015,Yoshida:2014}. Moreover, successful deposition of monolayer graphene onto the surface of these materials has been achieved \cite{Rizzo:2022,Rizzo:2020,Zhou:2019a,Zheng:2023,Rossi:2023,Leahy:2017,Altvater:2022,Boix-Constant:2021,Altvater:2021,Zhao:2017}. These advancements give us hope that our proposed tunneling junction could be realized in the near future. 
However, it is worth noting that these interfaces often exhibit significant charge transfer between graphene and the candidate quantum spin liquid due to the difference in the work functions of the materials, i.e., $\SI{0.5}{\electronvolt}$ \cite{Altvater:2021} and $\SI{1.5}{\electronvolt}$ \cite{Zhou:2019} for graphene on 1T-\ch{TaS2} and, respectively. In a dual-gated device, with independent control of the gates of the bottom and top graphene layers, we might hope to mitigate this charge transfer. This strategy seems particularly suitable for 1T-\ch{TaS2}, where graphene deposited on top of the bulk material gets hole doped by $\SI{0.1}{\electronvolt}$ \cite{Altvater:2021}.
Alternatively, an additional \ch{hBN} or TMD spacer layer between graphene and the QSL material, e.g., a monolayer on each side of the junction, could further reduce the doping to an acceptable level. The additional layer between the graphene and the candidate QSL material further allows to cap air-sensitive materials like 1T-\ch{TaS2} in the glove box and more easily transfer them in the quantum twisting microscope \cite{Inbar:2023}. 

In conclusion, we presented a proposal to detect the signatures of fractionalized excitations in two-dimensional materials via tunneling spectroscopy in a quantum twisting microscope. Importantly, this technique extends to the monolayer limit and offers both momentum and energy resolution, going beyond the capabilities of current experimental methods to probe magnetic materials. 

\begin{acknowledgments}
V.P. acknowledges the generous support by the Gordon and Betty Moore Foundation’s EPiQS Initiative, Grant GBMF8682. P.A.L. is grateful for the support by DOE office of Basic Sciences grant number DE-FG02- 03ER46076. S.I. is supported by the Leona M. and Harry B. Helmsley Charitable Trust grant, and the Rosa and Emilio Segre Research Award. G.R. expresses gratitude for the support by the Simons Foundation, the NSF DMR grant number 1839271 and the ARO MURI Grant No. W911NF-16-1-0361. G.R. and V.P. appreciate the support received from the Institute of Quantum Information and Matter. This work was completed at the Aspen Center for Physics, which is supported by the National Science Foundation grant PHY-2210452. 
\end{acknowledgments}

\begin{widetext}
\appendix

\section{Derivation of the tunneling current \label{app:one} 
} We here present the derivation of the tunneling current flowing through the junction. Our derivation parallels the one performed in Ref.~\onlinecite{Carrega:2020}. Our starting point is the tunneling Hamiltonian: 
\begin{equation}
	H_\tun=\frac{1}{\sqrt{N}}\sum_{\vec{k},\vec{q}}\sum_{\alpha,\beta,s,s'}\sum_{n=0}^2 T^{(n)}_{\alpha\beta} \left[\bar{\Gamma}_0\delta_{s,s'}\delta_{\vec{q},\Delta\vec{K}_n}+\bar{\Gamma}_1\vec{\sigma}_{ss'}\cdot \vec{s}_{\vec{q}_n}\right] \left(c^\dagger_{\vec{k},\alpha,s,\top}c^{\phantom{\dagger}}_{\vec{k}-\vec{q},\beta,s',\bot}e^{\mathrm{i}eV_\tun t} + c^\dagger_{\vec{k}-\vec{q},\alpha,s,\bot}c^{\phantom{\dagger}}_{\vec{k},\beta,s',\top}e^{-\mathrm{i}eV_\tun t}\right), 
\end{equation}
where the voltage bias across the tunneling junction is captured by a time-dependent tunneling process. This approach allows us to use equilibrium techniques to compute the tunneling current in linear response theory. The current operator is: 
\begin{equation}
	\begin{split}
		I_\tun &= \frac{\mathrm{i}e}{\hbar}\left[H_\tun,N_\top\right]\\&= -\frac{\mathrm{i}e}{\hbar\sqrt{N}} \sum_{\vec{k},\vec{q}}\sum_{\alpha,\beta,s,s'}\sum_{n=0}^2T_{\alpha\beta}^{(n)} \left[\bar{\Gamma}_0\delta_{s,s'}\delta_{\vec{q},\Delta\vec{K}_n}+\bar{\Gamma}_1\vec{\sigma}_{ss'}\cdot \vec{s}_{\vec{q}_n}\right]\left(c^\dagger_{\vec{k},\alpha,s,\top}c^{\phantom{\dagger}}_{\vec{k}-\vec{q},\beta,s',\bot}e^{\mathrm{i}eV_\tun t}-c^\dagger_{\vec{k}-\vec{q},\alpha,s,\bot}c^{\phantom{\dagger}}_{\vec{k},\beta,s',\top}e^{-\mathrm{i}eV_\tun t}\right). 
	\end{split}
\end{equation}
The current through the junction is then given by: 
\begin{equation}
	I= \frac{\mathrm{i}}{\hbar}\int_{-\infty}^{+\infty}\mathrm{d}t'\theta(t-t')\left\langle\left[H_\tun(t'),I_\tun(t)\right]\right\rangle= \frac{2e}{\hbar}\mathrm{Im}\chi_{AA}(eV_\tun), 
\end{equation}
with the time evolution generated by $H_0$. The operator $A$ is defined as: 
\begin{equation}\label{eq:appAOperator} 
	A(t)=\frac{1}{\sqrt{N}}\sum_{\vec{k},\vec{q}}\sum_{\alpha,\beta,s,s'}\sum_{n=0}^2 T_{\alpha\beta}^{(n)}\left[\bar{\Gamma}_0\delta_{s,s'}\delta_{\vec{q},\Delta\vec{K}_n}+\bar{\Gamma}_1\vec{\sigma}_{ss'}\cdot \vec{s}_{\vec{q}_n}(t)\right]c^\dagger_{\vec{k},\alpha,s,\top}(t)c^{\phantom{\dagger}}_{\vec{k}-\vec{q},\beta,s',\bot}(t), 
\end{equation}
and the response function $\chi_{AA}$ is: 
\begin{equation}
	\chi_{AA}(\omega)=-\frac{\mathrm{i}}{\hbar}\lim_{\eta\to 0^+}\int_0^\infty\mathrm{d}te^{\mathrm{i}(\omega+\mathrm{i}\eta) t/\hbar}\left\langle\left[A(t),A^\dagger(0)\right]\right\rangle. 
\end{equation}

The current can be easily evaluated via the Matsubara formalism. We separately compute the contribution stemming from elastic tunneling $\chi^{(0)}$ and the one from the spin-dependent scattering from the quantum spin liquid $\chi^{(2)}$. To this end, we define as $A_0$ the spin-independent part of the operator in Eq.~\eqref{eq:appAOperator} and as $A_2$ the spin-dependent one.

First, let us consider the elastic tunneling contribution: 
\begin{equation}\label{eq:appCurrentDirect} 
	\chi^{(0)}(\mathrm{i}\omega_n)=-\int_0^\beta\mathrm{d}\tau e^{\mathrm{i}\omega_n\tau}\langle\mathcal{T}A_0^{\phantom{\dagger}}(\tau)A_0^\dagger(0) \rangle=-\frac{2\bar{\Gamma}_0^2}{N}\sum_{\vec{k}}\sum_{n=0}^2\int_0^\beta\mathrm{d}\tau e^{\mathrm{i}\omega_n\tau}\,\mathrm{Tr}\,\left[\mathcal{G}^\bot(\vec{k}-\Delta\vec{K}_n,\tau)T^{(n)}\mathcal{G}^\top(\vec{k},-\tau)T^{(n)}\right], 
\end{equation}
where we assumed that the two graphene layers are uncorrelated in the absence of the tunneling Hamiltonian. The factor two comes from the spin multiplicity, and $\mathrm{Tr}$ acts on the orbital space. In Eq.~\eqref{eq:appCurrentDirect}, we introduced the electron Green's functions: 
\begin{align}
	\mathcal{G}_{\alpha,\alpha'}^\ell(\vec{k},-\tau)&=-\left\langle \mathcal{T} c^\dagger_{\vec{k},\alpha',s,\ell}(\tau)c^{\phantom{\dagger}}_{\vec{k},\alpha,s,\ell} \right\rangle,\\
	\mathcal{G}_{\alpha,\alpha'}^\ell(\vec{k},\tau)&=-\left\langle \mathcal{T} c^{\phantom{\dagger}}_{\vec{k},\alpha,s,\ell}(\tau)c^{\dagger}_{\vec{k},\alpha',s,\ell} \right\rangle. 
\end{align}
We conveniently parametrize the matrices $T^{(n)}$ as: 
\begin{equation}
	T^{(n)}=\tau^0+\cos{\left(\frac{2\pi n}{3}\right)}\tau^x-\sin{\left(\frac{2\pi n}{3}\right)}\tau^y, 
\end{equation}
and move from orbital to band space. Via a Fourier transform of the electronic Green's functions, we then obtain: 
\begin{equation}
	\chi^{(0)}(\mathrm{i}\omega_n)=-\frac{2\bar{\Gamma}_0^2}{N}\sum_{\vec{k},\lambda,\lambda'}\sum_{n=0}^2\left\lvert\bra{\lambda,\vec{k}}T^{(n)}\ket{\lambda',\vec{k}-\Delta\vec{K}_n} \right\rvert^2\frac{1}{\beta}\sum_{\nu_l}\mathcal{G}^\top_{\lambda}(\vec{k},\mathrm{i}\nu_l)\mathcal{G}^\bot_{\lambda'}(\vec{k}-\Delta\vec{K}_n,\mathrm{i}\nu_{l}+\mathrm{i}\omega_n), 
\end{equation}
with $\bra{\lambda,\vec{k}}T^{(n)}\ket{\lambda',\vec{k}-\Delta\vec{K}_n}$ the projection of the tunneling matrices on the eigenstates of band $\lambda$ and momentum $\vec{k}$ of the top graphene layer, and band $\lambda'$ and momentum $\vec{k}-\Delta\vec{K}_n$ for the bottom one, cf. Eq.~\eqref{eq:projectedT}. Lastly, we perform the Matsubara sum over the fermionic frequency $\nu_l$, perform an analytical continuation, and take the imaginary part to obtain Eq.~\eqref{eq:directTun}.

The derivation of the spin-dependent tunneling current proceeds in an analogous way. We need to compute: 
\begin{equation}\label{eq:appPartialChi2} 
	\begin{split}
		\chi^{(2)}(\mathrm{i}\omega_n)&=-\int_0^\beta\mathrm{d}\tau e^{\mathrm{i}\omega_n\tau}\langle\mathcal{T}A_2^{\phantom{\dagger}}(\tau)A_2^\dagger(0) \rangle\\&=-\frac{\bar{\Gamma}_1^2}{N}\sum_{\vec{k},\vec{k}'}\sum_{n=0}^2\int_0^\beta\mathrm{d}\tau e^{\mathrm{i}\omega_n\tau}\sum_\gamma\left\langle \mathcal{T} s^\gamma_{\vec{k}-\vec{k}'+\Delta\vec{K}_n}(\tau) s^\gamma_{-\vec{k}+\vec{k}'+\Delta\vec{K}_n}\right\rangle\,\mathrm{Tr}\,\left[\mathcal{G}^\bot(\vec{k}',\tau)T^{(n)}\mathcal{G}^\top(\vec{k},-\tau)T^{(n)}\right], 
	\end{split}
\end{equation}
where we assumed that, in the absence of the tunneling Hamiltonian, the quantum spin liquid and graphene layers are uncorrelated. The trace in Eq.~\eqref{eq:appPartialChi2} acts again on orbital space and we used that for Pauli matrices $\mathrm{tr}\,\left(\sigma^i\sigma^j\right)=\sum_{s,z}\sigma^i_{sz}\sigma^j_{zs}=\delta_{i,j}$. The Fourier transform to Matsubara frequency space leads to: 
\begin{equation}
	\chi^{(2)}(\mathrm{i}\omega_n)=\sum_{\vec{q}}\sum_{n=0}^2\frac{\bar{\Gamma}_1^2}{\beta}\sum_{\Omega_m}\mathcal{G}_\mathrm{s}(\vec{q}+\Delta\vec{K}_n,\mathrm{i}\omega_n+\mathrm{i}\Omega_m)\mathcal{G}_\tun(\vec{q},\mathrm{i}\Omega_m,n), 
\end{equation}
with $\vec{q}=\vec{k}-\vec{k}'$ and 
\begin{align}
	&\mathcal{G}_\tun(\vec{q},\mathrm{i}\Omega_m,n)=\frac{1}{N}\sum_{\vec{k},\lambda,\lambda'}\left\lvert\bra{\lambda,\vec{k}}T^{(n)}\ket{\lambda',\vec{k}-\vec{q}} \right\rvert^2 \frac{1}{\beta}\sum_{\nu_l}\mathcal{G}^\top_{\lambda}(\vec{k},\mathrm{i}\nu_l)\mathcal{G}^\bot_{\lambda'}(\vec{k}-\vec{q},\mathrm{i}\nu_{l}-\mathrm{i}\Omega_m),\\
	&\mathcal{G}_\mathrm{s}(\vec{q}+\Delta\vec{K}_n,\mathrm{i}\omega_n+\mathrm{i}\Omega_m)=-\int_0^\beta\mathrm{d}\tau e^{\mathrm{i}(\omega_n+\Omega_m)\tau}\left\langle \mathcal{T} s^\gamma_{\vec{k}-\vec{k}'+\Delta\vec{K}_n}(\tau) s^\gamma_{-\vec{k}+\vec{k}'+\Delta\vec{K}_n}\right\rangle. 
\end{align}
Lastly, we perform the Matsubara sums, perform an analytical continuation and take the imaginary part to obtain Eq.~\eqref{eq:final}.

In all generality, one would be an additional contribution to the total current linear in $\bar{\Gamma}_1$. This additional term, however, vanishes for spin non-polarized leads as it is proportional to $\langle \vec{s}\rangle$. This last term is given by: 
\begin{equation}
	\begin{split}
		\chi^{(1)}(\mathrm{i}\omega_n)&=-\int_0^\beta\mathrm{d}\tau e^{\mathrm{i}\omega_n\tau}\langle\mathcal{T}A_2^{\phantom{\dagger}}(\tau)A_0^\dagger(0) \rangle\\&=-\frac{\bar{\Gamma}_0\bar{\Gamma}_1}{N}\sum_\gamma\sum_{\vec{k},\vec{q}}\sum_{n,s}\sigma_{ss}^\gamma\left\langle \mathcal{T} s^\gamma_{\vec{q}+\Delta\vec{K}_n}(\tau) \right\rangle\mathrm{Tr}\,\left[\mathcal{G}^\bot(\vec{k}+\Delta\vec{K}_n,\tau)T^{(n)}\mathcal{G}^\top(\vec{k},-\tau)T^{(n)}\right]. 
	\end{split}
\end{equation}
In the absence of spin polarization in the leads, the term $\sum_\mathrm{s} \sigma_{ss}^\gamma$ results in a vanishing contribution.

\section{Contribution of the elastic tunneling to the total current \label{grapheneeh} 
} We here derive the contribution to the total current stemming from elastic tunneling. We restrict ourselves to the limit $\mu>eV_\tun$ and consider electron-doped graphene layers. Therefore, we consider tunneling events involving only the conduction bands of the two layers, i.e., $\lambda=\lambda'=+1$. First, from Eq.~\eqref{eq:projectedT}, we note that: 
\begin{equation}\label{eq:scatteringAmplitude} 
	\lvert T^{(n)}_{+1,+1}\rvert^2 = \left[1+\cos\left(\frac{2\pi n}{3}+\phi_{\vec{k}}\right)\right]\left[1+\cos\left(\frac{2\pi n}{3}+\phi_{\vec{k}-\vec{q}}\right)\right]. 
\end{equation}
We then have: 
\begin{equation}
	\begin{split}
		I^{(0)}_{T=0}(eV_\tun)&=-\frac{\sqrt{3}ea_\mathrm{g}^2\bar{\Gamma}_0^2}{2\pi\hbar}\sum_n\int_0^{eV_\tun}\mathrm{d}\epsilon\int\mathrm{d}^2\vec{k}\left[1+\cos\left(\frac{2\pi n}{3}+\phi_{\vec{k}}\right)\right]\left[1+\cos\left(\frac{2\pi n}{3}+\phi_{\vec{k}-\Delta\vec{K}_n}\right)\right]\\&\quad\times\delta\left(\epsilon-eV_\tun-\hbar v_\mathrm{F}\lvert\vec{k}\rvert+\mu\right)\delta\left(\epsilon-\hbar v_\mathrm{F}\lvert\vec{k}-\Delta\vec{K}_n\rvert+\mu\right)\\
		&=-\frac{\sqrt{3}ea_\mathrm{g}^2\bar{\Gamma}_0^2}{2\pi\hbar^2 v_\mathrm{F}}\sum_n\int\mathrm{d}^2\vec{k}\left[1+\cos\left(\frac{2\pi n}{3}+\phi_{\vec{k}}\right)\right]\left[1+\cos\left(\frac{2\pi n}{3}+\phi_{\vec{k}-\Delta\vec{K}_n}\right)\right]\\&\quad\times\delta\left(\frac{eV_\tun}{\hbar v_\mathrm{F}}+\lvert\vec{k}\rvert-\lvert\vec{k}-\Delta\vec{K}_n\rvert\right)\Theta\left(eV_\tun-\hbar v_\mathrm{F}\lvert\vec{k}-\Delta\vec{K}_n\rvert+\mu\right), 
	\end{split}
\end{equation}
where we converted the sum over $\vec{k}$ in an integral. To perform the integral, we introduce the new variable $p=\lvert\vec{k}-\Delta\vec{K}_n\rvert$ and replace the integral over $\phi_{\vec{k}}$ in $\int\mathrm{d}^2\vec{k}=\int_0^\infty \mathrm{d}k k\int_0^{2\pi}\mathrm{d}\phi_{\vec{k}}$ via \cite{Agarwal:2020}: 
\begin{equation}
	\int \mathrm{d}^2\vec{k}=\int_0^\infty \mathrm{d}k\int_0^{2\pi}\mathrm{d}\phi_{\vec{k}} \int_0^\infty\mathrm{d}p\, \frac{p}{\Delta K_n}\delta\left(\frac{p^2-\Delta K_n^2-k^2}{2k\Delta K_n}+\cos\phi_{\vec{k}}\right). 
\end{equation}

We also express the cosine and sine of the angles $\phi_{\vec{k}}$ and $\phi_{\vec{k}+\Delta\vec{K}_n}$ in Eq.~\eqref{eq:scatteringAmplitude} in terms of the new variables $k$ and $p$: 
\begin{eqnarray}
	&\cos\phi_{\vec{k}} = \frac{\Delta K_n^2+k^2-p^2}{2\Delta K_nk},\\
	& \cos\phi_{\vec{k}-\Delta\vec{K}_n}=\frac{k^2-\Delta K_n^2-p^2}{2\Delta K_np},\\
	& \sin\phi_{\vec{k}}=\frac{\sqrt{4\Delta K_n^2k^2-(\Delta K_n^2+k^2-p^2)^2}}{2\Delta K_nk},\\
	& \sin\phi_{\vec{k}-\Delta\vec{K}_n}=\frac{\sqrt{4\Delta K_n^2p^2-(k^2-\Delta K_n^2-p^2)^2}}{2\Delta K_np}. 
\end{eqnarray}

We can now carry out the integral over $\phi_{\vec{k}}$: 
\begin{equation}
	\begin{split}
		\int_0^{2\pi}\mathrm{d}\phi_{\vec{k}}\,\delta\left(\frac{p^2-\Delta K_n^2-k^2}{2k\Delta K_n}+\cos\phi_{\vec{k}}\right) &= \frac{1}{2\pi}\int_{-\infty}^{+\infty}\mathrm{d}x \int_0^{2\pi}\mathrm{d}\phi_{\vec{k}} e^{\mathrm{i}x\left(\frac{p^2-\Delta K_n^2-k^2}{2k\Delta K_n}+\cos\phi_{\vec{k}}\right)}\\
		&= \int_{-\infty}^{+\infty} \mathrm{d}x e^{\mathrm{i}x\frac{p^2-\Delta K_n^2-k^2}{2k\Delta K_n}}J_0(\lvert x\rvert)= \frac{4k\Delta K_n}{\sqrt{\left[(p+k)^2-\Delta K_n^2\right]\left[\Delta K_n^2-(p-k)^2\right]}}, 
	\end{split}
\end{equation}
where $J_0(x)$ is the zero Bessel's function of the first kind. 

With these replacements, and carrying out the sum over $n$, we obtain: 
\begin{equation}
	\begin{split}
		I^{(0)}_{T=0}(eV_\tun)&=-\frac{3\sqrt{3}ea_\mathrm{g}^2\bar{\Gamma}_0^2}{2\pi\hbar^2 v_\mathrm{F}}\int_0^{\infty}\mathrm{d}k\int_0^\infty\mathrm{d}p\,\delta\left(\frac{eV_\tun}{\hbar v_\mathrm{F}}+k-p\right)\Theta\left(eV_\tun-\hbar v_\mathrm{F}p+\mu\right)\\&\quad \times\frac{4kp}{\sqrt{\left[(k+p)^2-\Delta K_n^2\right]\left[\Delta K_n^2-(p-k)^2\right]}}\left\{1+\frac{k^2+p^2-\Delta K_n^2}{4pk}\right\}, 
	\end{split}
\end{equation}

The remaining integrals can now be readily solved to obtain: 
\begin{equation}
	\begin{split}
		I^{(0)}_{T=0}(eV_\tun)&=-\frac{3\sqrt{3}ea^2_\mathrm{g}\bar{\Gamma}_0^2}{16\pi\hbar^3 v_\mathrm{F}^2\sqrt{\hbar^2v_\mathrm{F}^2\Delta K_n^2-eV_\tun^2}}\Theta\left[\hbar v_\mathrm{F}\Delta K_n-eV_\tun\right]\Theta\left[eV_\tun+2\mu-\hbar v_\mathrm{F}\Delta K_n\right]\\&\quad\times\Bigg[3(eV_\tun+2\mu)\sqrt{(eV_\tun+2\mu)^2-\hbar^2 v_\mathrm{F}^2\Delta K_n^2} \\&\quad \quad\quad-\left(\hbar^2 v_\mathrm{F}^2\Delta K^2_n+2eV_\tun^2\right)\log\left(\frac{eV_\tun+2\mu}{\hbar v_\mathrm{F}\Delta K_n}+\sqrt{\frac{(eV_\tun+2\mu)^2}{\hbar^2 v_\mathrm{F}^2\Delta K_n^2}-1}\right) \Bigg], 
	\end{split}
\end{equation}
As stressed in the main text, the elastic tunneling contributes to the total current, and hence to the IETS signal, only if $eV_\tun<\hbar v_\mathrm{F}\lvert\Delta \vec{K}_n\rvert<eV_\tun+2\mu$.

\section{Derivation of the spin structure factor \label{app:three} 
} As discussed in Sec.~\ref{sec:quantum spin liquid}, we represent the physical spins in terms of Abrikosov fermions. We choose fermions rather than Schwinger bosons as they allow us to easily treat gapless quantum spin liquid without worrying about the condensation of the spinons. We define 
\begin{equation}
	\vec{s}(\vec{r})=\frac{1}{2}\sum_{s,s'}f^\dagger_{\vec{r},s}\vec{\sigma}_{s s'}f^{\phantom{\dagger}}_{\vec{r}\beta}. 
\end{equation}
 where $f_{\vec{r},s}$ annihilates a spinon at location $\vec{r}$ with spin $s$. Imposing constraints to avoid unphysical states, such as empty or doubly occupied sites, i.e., $\sum_s f^\dagger_{\vec{r}s}f^{\phantom{\dagger}}_{\vec{r}s}=1$ and $f_{is}f_{is'}\epsilon_{ss'}=0$, introduces additional gauge degrees of freedom.
At the mean-field level, we neglect gauge fluctuations and enforce the constraints only on average. By introducing auxiliary fields to decouple the quartic term in the spinon operators, we obtain the quadratic Hamiltonian for the spinons that serves as the basis for further analysis:
\begin{equation}\label{eq:mfSpinon}
H_\qsl=\sum_{\vec{r}_i\vec{r}_j,ss'}\left(t_{\vec{r}_i\vec{r}_j}f^\dagger_{is} f^{\phantom{\dagger}}_{js}+\Delta_{\vec{r}_i\vec{r}_j}f^\dagger_{is} f^\dagger_{js'}+\mathrm{H.c.}\right).
\end{equation}
Here, $\Delta\neq 0$ describes a $\mathbb{Z}_2$ QSL, whereas, a $U(1)$ QSL requires $\Delta=0$. The chemical potential, included in $t_{\vec{r}_i\vec{r}_j}$, enforces the half-filling condition of the spinon bands.

We now derive the spin structure factor expressed in terms of the spinon operators, as shown in Eq.~\eqref{eq:finalquantum spin liquids}. We will consider the possibility of multiple sublattices in the unit cell at locations $\vec{r}_i=\vec{R}_i+\vec{\delta}_\alpha$, where $\vec{R}_i$ is the unit cell center and $\vec{\delta}_\alpha$ the position of the sublattice inside the unit cell. 
\begin{equation}\label{eq:spinonSpinStructureSupp} 
	\begin{split}
		\mathcal{G}_\mathrm{s}(\vec{q},\mathrm{i} \omega_n)&=\frac{1}{N}\sum_\gamma\sum_{\alpha,\beta}\int_0^\beta\mathrm{d}\tau e^{\mathrm{i}\tau\omega_n}\sum_{ij}e^{i\vec{q}\cdot(\vec{r}_i-\vec{r}_j)}\langle \mathcal{T}s^\gamma_{i\alpha}(\tau)s^\gamma_{j\beta}\rangle \\&=\frac{1}{4N}\sum_\gamma\sum_{\substack{ss'\\
		zz'}}\sum_{\alpha,\beta}\sigma^\gamma_{ss'}\sigma^\gamma_{zz'}\int_0^\beta\mathrm{d}\tau e^{\mathrm{i}\tau\omega_n}\sum_{ij}e^{\mathrm{i}\vec{q}\cdot(\vec{r}_i-\vec{r}_j)}\langle \mathcal{T}f^\dagger_{\alpha is}(\tau) f^{\phantom{\dagger}}_{\alpha is'}(\tau) f^\dagger_{\beta jz}f^{\phantom{\dagger}}_{\beta jz'} \rangle\\
		&= \frac{1}{4N}\sum_{ss'}\sum_{\alpha,\beta}\int_0^\beta\mathrm{d}\tau e^{\mathrm{i}\tau\omega_n}\sum_{ij}e^{\mathrm{i}\vec{q}\cdot(\vec{r}_i-\vec{r}_j)}\left[2\langle\mathcal{T} f^\dagger_{\alpha is}(\tau) f^{\phantom{\dagger}}_{\alpha is'}(\tau) f^\dagger_{\beta js'}f^{\phantom{\dagger}}_{\beta js}\rangle - \langle\mathcal{T} f^\dagger_{\alpha is}(\tau) f^{\phantom{\dagger}}_{\alpha is}(\tau) f^\dagger_{\beta js'}f^{\phantom{\dagger}}_{\beta js'}\rangle\right]\\
		&=\frac{1}{4N}\sum_{\vec{k}\vec{k}'}\sum_{ss'}\sum_{\alpha,\beta}\int_0^\beta\mathrm{d}\tau e^{\mathrm{i}\tau\omega_n}e^{\mathrm{i}\vec{q}\cdot(\vec{\delta}_\alpha-\vec{\delta}_\beta)}\Big[2\langle\mathcal{T} f^\dagger_{\alpha\vec{k}s}(\tau) f^{\phantom{\dagger}}_{\alpha\vec{k}+\vec{q}s'}(\tau) f^\dagger_{\beta\vec{k}'s'}f^{\phantom{\dagger}}_{\beta\vec{k}'-\vec{q}s}\rangle \\&\quad\quad\quad\quad\quad\quad\quad\quad\quad\quad\quad\quad\quad\quad\quad\quad\quad\quad\quad- \langle\mathcal{T} f^\dagger_{\alpha\vec{k}s}(\tau) f^{\phantom{\dagger}}_{\alpha\vec{k}+\vec{q}s}(\tau) f^\dagger_{\beta\vec{k}'s'}f^{\phantom{\dagger}}_{\beta\vec{k}'-\vec{q}s'}\rangle\Big]. 
	\end{split}
\end{equation}
where, in the second line, we used that $\sum_\gamma \sigma_{ss'}^\gamma\sigma^\gamma_{zz'}=2\delta_{s,z'}\delta_{z,s'}-\delta_{s,s'}\delta_{z,z'}$.

We restrict ourselves to the case of a U(1) quantum spin liquid. As a first step, we move from orbital to band basis: 
\begin{equation}
	f_{n\vec{k}s}=\sum_\alpha U_{\alpha ns}f_{\alpha \vec{k}s}, 
\end{equation}
where $U_{\alpha ns}$ is the $n^\mathrm{th}$ eigenvector of the spinon mean-field Hamiltonian at momentum $\vec{k}$ with spin $s$ at sublattice $\alpha$. 
\begin{equation}
	\begin{split}
		\mathcal{G}_\mathrm{s}(\vec{q},\mathrm{i}\omega_n)&=\frac{1}{4N}\sum_{\vec{k}\vec{k}'}\sum_{\alpha\beta}\sum_{s,s'}\sum_{nlqt}\int_0^\beta\mathrm{d}\tau e^{\mathrm{i}\tau\omega_n}e^{\mathrm{i}\vec{q}\cdot(\vec{\delta}_\alpha-\vec{\delta}_\beta)}\\
		&\quad\times \Bigg[2U^*_{\alpha ns}(\vec{k})U_{\alpha ls'}(\vec{k}+\vec{q})U^*_{\beta qs'}(\vec{k}')U_{\beta ts}(\vec{k}'-\vec{q})\langle \mathcal{T}f^\dagger_{n\vec{k}s}(\tau) f^{\phantom{\dagger}}_{l\vec{k}+\vec{q}s'}(\tau) f^\dagger_{q\vec{k}'s'}f^{\phantom{\dagger}}_{t\vec{k}'-\vec{q}s}\rangle \\
		&\quad\quad-U^*_{\alpha ns}(\vec{k})U_{\alpha ls}(\vec{k}+\vec{q})U^*_{\beta qs'}(\vec{k}')U_{\beta ts'}(\vec{k}'-\vec{q}) \langle\mathcal{T} f^\dagger_{n\vec{k}s}(\tau) f^{\phantom{\dagger}}_{l\vec{k}+\vec{q}s}(\tau) f^\dagger_{q\vec{k}'s'}f^{\phantom{\dagger}}_{t\vec{k}'-\vec{q}s'}\rangle\Bigg]\\
		&=\frac{1}{4N}\sum_{\vec{k}\vec{k}'}\sum_{\alpha\beta}\sum_{s,s'}\sum_{nlqt}\int_0^\beta\mathrm{d}\tau e^{\mathrm{i}\tau\omega_n}e^{\mathrm{i}\vec{q}\cdot(\vec{\delta}_\alpha-\vec{\delta}_\beta)}U^*_{\alpha ns}(\vec{k})U_{\alpha ls'}(\vec{k}+\vec{q})U^*_{\beta qs'}(\vec{k}')U_{\beta ts}(\vec{k}'-\vec{q})\\
		&\quad \times \left[2 G^n_{ss}(\vec{k},-\tau)G^l_{s's'}(\vec{k}+\vec{q},\tau)\delta_{nt}\delta_{lq}\delta_{\vec{k},\vec{k}'-\vec{q}}-G^n_{ss}(\vec{k},-\tau)G^l_{s's'}(\vec{k}+\vec{q},\tau)\delta_{ss'}\delta_{nt}\delta_{lq}\delta_{\vec{k},\vec{k}'-\vec{q}}\right]\\
		&=\frac{1}{4N}\sum_{\vec{k}}\sum_{ss'}\sum_{nl}\int_0^\beta\mathrm{d}\tau e^{\mathrm{i}\tau\omega_n}g_{ss'}(\vec{k},\vec{q},n,l) (2-\delta_{ss'})G^n_{ss}(\vec{k},-\tau)G^l_{s's'}(\vec{k}+\vec{q},\tau). 
	\end{split}
\end{equation}

We introduced the spinons Green's functions: 
\begin{equation}
	\begin{split}
		G^n_{ss}(\vec{k},\tau)&=\langle\mathcal{T} f^\pdag_{ns\vec{k}}(\tau)f^\dagger_{ns\vec{k}}\rangle,\\
		G^n_{ss}(\vec{k},-\tau)&=\langle\mathcal{T} f^\dagger_{ns\vec{k}}(\tau)f^\pdag_{ns\vec{k}}\rangle, 
	\end{split}
\end{equation}
and the function $g$ that accounts for the overlap of the eigenstates' wavefunctions: 
\begin{equation}
	g_{ss'}(\vec{k},\vec{q},n,l)=\left\vert\sum_{\alpha} e^{\mathrm{i}\vec{q}\cdot\vec{\delta}_\alpha}U^*_{\alpha ns}(\vec{k})U_{\alpha ls'}(\vec{k}+\vec{q}) \right\vert^2. 
\end{equation}

With the fermionic Matsubara frequencies $\nu_l$ and $\Omega_l$, we can rewrite the spinon Green's function in frequency space: 
\begin{equation}
	\begin{split}
		G^n_{ss}(\vec{k}+\vec{q},\tau)&=\frac{1}{\beta}\sum_{\Omega_l}e^{\mathrm{i}\Omega_l\tau}G^n_{ss}(\vec{k}+\vec{q},\mathrm{i}\Omega_l),\\
		G^n_{ss}(\vec{k},-\tau)&=\frac{1}{\beta}\sum_{\nu_l}e^{-\mathrm{i}\nu_l\tau}G^n_{ss}(\vec{k},\mathrm{i}\nu_l). 
	\end{split}
\end{equation}
We then obtain: 
\begin{equation}
	\mathcal{G}_\mathrm{s}(\vec{q},\mathrm{i}\omega_n)=\frac{1}{4N}\sum_{\vec{k}}\sum_{ss'}\sum_{nl}\frac{1}{\beta}\sum_{\Omega_l}g_{ss'}(\vec{k},\vec{q},n,l) (2-\delta_{ss'})G^n_{ss}(\vec{k},\mathrm{i}\Omega_l+\mathrm{i}\omega_n)G^l_{s's'}(\vec{k}+\vec{q},\mathrm{i}\Omega_l). 
\end{equation}
Note that $\mathrm{i}\omega_n=\mathrm{i}\nu_l-\mathrm{i}\Omega_l$ is a bosonic Matsubara frequency. We can finally perform the Matsubara sum over $\Omega_l$ and obtain 
\begin{equation}
	\mathcal{G}_\mathrm{s}(\vec{q},\mathrm{i}\omega_n) = -\frac{1}{4N} \sum_{ss'}\sum_{\vec{k}}\sum_{nl} g_{ss'}(\vec{k},\vec{q},n,l) (2-\delta_{ss'})\frac{n_\mathrm{F}(\xi^s_{n\vec{k}})-n_\mathrm{F}(\xi^{s'}_{l\vec{k}+\vec{q}})}{\mathrm{i}\omega_n-\xi^s_{n\vec{k}}+\xi^{s'}_{l\vec{k}+\vec{q}}}. 
\end{equation}
Finally, performing an analytical continuation and taking the imaginary part, we reach: 
\begin{equation}\label{eq:finalquantum spin liquidsApp} 
	\mathcal{S}(\vec{q},\omega) = \frac{1}{4N} \sum_{ss'}\sum_{\vec{k}}\sum_{nl} g_{ss'}(\vec{k},\vec{q},n,l) (2-\delta_{ss'})\left[n_\mathrm{F}(\xi^{s}_{n\vec{k}})-n_\mathrm{F}(\xi^{s'}_{l\vec{k}+\vec{q}})\right]\delta\left(\omega-\xi^s_{n\vec{k}}+\xi^{s'}_{l\vec{k}+\vec{q}}\right), 
\end{equation}
which corresponds to Eq.~\eqref{eq:finalquantum spin liquids}.

\subsection{Dirac quantum spin liquid} We can first study a gapless U(1) quantum spin liquid where spinons have a massless Dirac dispersion: a Dirac quantum spin liquid. The low energy dispersion is given by $\xi_{n\vec{k}}=\hbar v_\mathrm{s}n\lvert\vec{k}\rvert$, with $n=\pm$ and $v_\mathrm{s}$ the spinon Fermi velocity.

The computation of spin structure factor of a Dirac quantum spin liquid is analogous to the derivation of the particle-hole spectral function of graphene presented in Sec.~\ref{grapheneeh}, without the additional complication of the scattering matrix of the graphene bilayer's structure.

We readily obtain: 
\begin{equation}
	\mathcal{S}(\vec{q},\omega) = \frac{3\Omega}{2}\int \frac{\mathrm{d}^2\vec{k}}{(2\pi)^2}\delta\left(\omega-\hbar v_\mathrm{s}\lvert\vec{k}\rvert-\hbar v_\mathrm{s}\lvert\vec{k}+\vec{q}\rvert\right)=\frac{3\Omega}{16\pi\hbar^2 v_\mathrm{s}^2}\frac{\omega^2-\hbar^2v_\mathrm{s}^2 q^2/2}{\sqrt{\omega^2-\hbar^2 v_\mathrm{s}^2 q^2}}\Theta\left(\omega-\hbar v_\mathrm{s}q\right), 
\end{equation}
where $\Omega$ is the ares of the QSL's unit cell.

\subsection{Chiral quantum spin liquid} To capture a U(1) chiral spin liquid, we consider a spinon model with a gap in the spectrum. For simplicity, we assume that the spinon dispersion has a minimum at $\vec{k}=0$ for the conduction band and a maximum at the same location for the valence band. The gap at $\vec{k}=0$ is $\Delta_\mathrm{s}$. We consider the dispersion $\xi_{n\vec{k}}=n\left(\frac{\hbar^2\vec{k}^2}{2m_\mathrm{s}}+\frac{\Delta_\mathrm{s}}{2}\right)$, with $n=\pm$ and $m_\mathrm{s}$ the spinon mass. At zero temperature, we need to consider exclusively transition from the valence to the conduction band, i.e., $n=+$ and $l=-$ in Eq.~\eqref{eq:finalquantum spin liquidsApp}.

The integral to compute is 
\begin{equation}
	\mathcal{S}(\vec{q},\omega) = \frac{3\Omega}{2}\int \frac{\mathrm{d}^2\vec{k}}{(2\pi)^2}\delta\left(\omega-\Delta_\mathrm{s}-\frac{\hbar^2k^2}{2m_\mathrm{s}}-\frac{\hbar^2(\vec{k}+\vec{q})^2}{2m_\mathrm{s}}\right). 
\end{equation}
It can be readily solved: 
\begin{equation}
	\begin{split}
		\mathcal{S}(\vec{q},\omega) &= \frac{3\Omega}{2}\int \frac{\mathrm{d}^2\vec{k}}{(2\pi)^2}\delta\left(\omega-\Delta-\frac{\hbar^2k^2}{m_\mathrm{s}}-\frac{\hbar^2q^2}{2m_\mathrm{s}}-\frac{\hbar^2kq\cos\phi_{\vec{k}}}{m_\mathrm{s}}\right)\\
		& = \frac{3\Omega}{2}\int \frac{\mathrm{d}^2\vec{k}}{(2\pi)^2}\delta\left(\omega-\Delta-\frac{\hbar^2 q^2}{4m_\mathrm{s}}-\frac{\hbar^2( \vec{k}-\vec{q}/2)^2}{m_\mathrm{s}}\right)\\
		& = \frac{3m_\mathrm{s}\Omega}{8\pi^2}\int_0^{\infty}\mathrm{d}p\int_0^{2\pi}\mathrm{d}\phi_{\vec{p}}\,p\delta\left[m_\mathrm{s}\left(\omega-\Delta-\frac{\hbar^2q^2}{4m_\mathrm{s}}\right)-\hbar^2p^2\right]\\
		& = \frac{3m_\mathrm{s}\Omega}{8\pi\hbar^2}\Theta\left(\omega-\Delta-\frac{\hbar^2q^2}{4m_\mathrm{s}}\right). 
	\end{split}
\end{equation}
A completely analogous result captures the zero-temperature low-energy and small-momentum spin structure factor of a gapped $\mathbb{Z}_2$ spin liquid. 

\subsection{Spinon Fermi surface} The calculation of the spin structure factor in the presence of a spinon Fermi surface is analogous to the calculation of the dynamical spin structure factor of a neutral Fermi gas \cite{Nozieres:1999}. 

For the spinons, we assume a simple parabolic dispersion with mass $m_\mathrm{s}$ and Fermi momentum $k_\mathrm{s,F}$. If we restrict to small momentum $\vec{q}$, the states contributing to the dynamical spin structure factor $\mathcal{S}(\vec{q},\omega)$ are those in the crescent defined by $\lvert \vec{k}\rvert<k_\mathrm{s,F}$ and $\lvert \vec{k}+\vec{q}\rvert>k_\mathrm{s,F}$. We have: 
\begin{equation}
	\begin{split}
		\mathcal{S}(\vec{q},\omega) &= \frac{3}{2}\frac{1}{2 k_\mathrm{s,F}q}\int_{\substack{k<k_\mathrm{s,F}\\
		\lvert \vec{k}+\vec{q}\rvert>k_\mathrm{s,F}}} \frac{\mathrm{d}^2\vec{k}}{(2\pi)^2}\delta\left(\omega+\frac{\hbar^2 k^2}{2m_\mathrm{s}}-\frac{\hbar^2 (\vec{k}+\vec{q})^2}{2m_\mathrm{s}}\right) \\
		& = \frac{3}{2}\frac{1}{2 k_\mathrm{s,F}q}\int_{\substack{k<k_\mathrm{s,F}\\
		\lvert \vec{k}+\vec{q}\rvert>k_\mathrm{s,F}}} \frac{\mathrm{d}^2\vec{k}}{(2\pi)^2}\delta\left(\omega-\frac{\hbar^2 q^2}{2m_\mathrm{s}}-\frac{\hbar^2 qk\cos\phi_{\vec{k}}}{m_\mathrm{s}}\right), 
	\end{split}
\end{equation}
where $\phi_{\vec{k}}$ is the angle between $\vec{k}$ and $\vec{q}$.

For small $\vec{q}$, we can express the number of state in the crescent between $\phi_{\vec{k}}$ and $\phi_{\vec{k}}+\mathrm{d}\phi_{\vec{k}}$ as: 
\begin{equation}
	\frac{\mathrm{d}^2\vec{k}}{(2\pi^2)} = \frac{k_\mathrm{s,F}q\cos\phi_{\vec{k}}\,\mathrm{d}\phi_{\vec{k}}}{(2\pi)^2}, 
\end{equation}
where we approximated $\mathrm{d}k=q\cos\phi_{\vec{k}}$. We finally obtain: 
\begin{equation}
	\begin{split}
		\mathcal{S}(\vec{q},\omega)&=\frac{3}{16\pi^2 \hbar q v_\mathrm{s}} \int_{-\pi/2}^{\pi/2}\mathrm{d}\phi_{\vec{k}} \cos\phi_{\vec{k}}\, \delta\left(\frac{\omega}{\hbar qv_\mathrm{s}}-\cos\phi_{\vec{k}}\right)\\
		&=\frac{3}{8\pi^2 \hbar qv_\mathrm{s}}\frac{\omega}{\sqrt{\hbar^2q^2v_\mathrm{s}^2-\omega^2}}\Theta\left(\hbar qv_\mathrm{s}-\omega\right), 
	\end{split}
\end{equation}
where the Fermi velocity is defined as $v_{\mathrm{s}}=\hbar k_\mathrm{s,F}/m_\mathrm{s}$, and we neglected the contribution $\hbar^2q^2/2m_\mathrm{s}$ in the delta function. This expression corresponds to Eq.~\eqref{eq:kinematicquantum spin liquid} of the main text. 
\end{widetext}
\bibliography{refs} 

\clearpage

\end{document}